\documentclass[aps, prc, amsmath, floats,floatfix, superscriptaddress,twocolumn]{revtex4-2}
\usepackage{graphicx}
\usepackage{epsfig}
\usepackage{hyperref}
\usepackage{ulem}
\usepackage{morefloats}
\usepackage{rotating}
\usepackage{multirow}
\usepackage{booktabs}
\usepackage{multirow}
\usepackage{natbib}
\usepackage{dcolumn}
\usepackage{bm}
\usepackage[utf8]{inputenc}
\usepackage[usenames,dvipsnames]{color}
\usepackage{amsmath,amssymb,bbold}

\usepackage{xcolor}
\newcommand{\be}{\begin{eqnarray}}
\newcommand{\ee}{\end{eqnarray}}

\begin{document}
\title{Constraining the high-density behavior of nuclear symmetry energy with direct Urca processes
}
\author{Olfa Boukari} 
\email{olfa.boukari@isepbg.ucar.tn}
\affiliation{ISEPBG-Soukra, University of Carthage, Avenue de la République BP 77-1054 Amilcar, Tunisia}
\author{Tuhin Malik}
\email{tm@uc.pt}
\affiliation{CFisUC, Department of Physics, University of Coimbra, 3004-516 Coimbra, Portugal}
\author{Aziz Rabhi} 
\email{rabhi@fis.uc.pt}
\affiliation{IPEST La Marsa, University of Carthage, Avenue de la République BP 77-1054 Amilcar, Tunisia}
\affiliation{CFisUC, Department of Physics, University of Coimbra, 3004-516 Coimbra, Portugal}
\author{Constan\c ca Provid\^encia}
\email{cp@uc.pt}
\affiliation{CFisUC, Department of Physics, University of Coimbra, 3004-516 Coimbra, Portugal}
\begin{abstract}
The density dependence of the symmetry energy in relativistic mean-field models with density dependent couplings is discussed in terms of the possible opening of nucleonic direct Urca processes inside neutron stars, which induce a very rapid cooling of the star. The modification of the parametrization of the isospin channel of two models, DD2 and DDMEX, keeping the same isoscalar properties is considered and the implications are discussed.  Within the models discussed it is not possible the onset of nucleonic direct Urca processes in stars with a mass below $\sim1.6\,M_\odot$ if chiral effective field theory constraints for neutron matter are imposed. A Bayesian inference calculation confirms the low probability that nucleonic direct Urca processes occur inside stars with masses below 1.8$M_\odot$, considering the isoscalar channel of the equation of state described by DD2 or DDMEX and the same symmetry energy at saturation. The lowest masses allowing direct Urca processes are associated with a slope of the symmetry energy above $60$ MeV and most likely a positive symmetry energy incompressibility. It is shown that the parametrization of the isospin channel proposed destroys the correlation between symmetry energy slope and incompressibility previously identified in several works.
\end{abstract}

\maketitle

\section{Introduction}

In the past decade, multi-messenger astronomy has made significant progress, particularly with the recent detection of gravitational waves (GW) by the LIGO  and Virgo Collaboration (LVC). Notably, they detected GW originating from two distinct binary neutron star (BNS) mergers. The first event, known as GW170817~\citep{abbott2017a, abbott2017b, abbott2018}, involved a BNS merger with a total mass of 2.7M$_\odot$ and mass components ranging between 1.17M$_\odot$ and 1.6M$_\odot$. Similarly, the subsequent detection of the GW190425 merger, as detailed in reference  \citep{abbott2020a}, further expanded our knowledge. This event featured a total mass of $3.4^{+0.3}_{-0.1}\mathrm{M}_\odot$, with mass components spanning from 1.12M$_\odot$ to 2.52M$_\odot$. These observations provide valuable insights into high-density stellar matter \cite{Miller2021,Annala2022,Biswas2022}, and offer new constraints for understanding the Equation of State (EoS) of stellar matter \cite{Lattimer12ARNPS,Oertel2017,Tolos2020,Hebeler2021}.

The ambitious goal of the astrophysics community is to combine these multi-messenger results to constrain the EoS of neutron stars (NSs)\cite{Miller2021,Raaijmakers_2021,JieLiJ21,Legred21,Pang21,TangSP21,Annala2022,Biswas2022}, and infer the NS composition.
The nuclear symmetry energy constitutes an essential ingredient to explore the macroscopic properties of NSs, and should satisfy theoretical constraints such as those obtained from chiral effective field theory (chEFT) calculations for pure neutron matter \cite{Hebeler2013}. It should also be consistent with experimental and observational data. 


Thanks to the considerable efforts deployed in astrophysics and nuclear physics over the last two decades, significant progress has been made in determining the symmetry energy  $E_{\textrm{sym}}(\rho_{B})$, in particular around and below the saturation density of nuclear matter $\rho_0$.

The symmetry energy $E_{\rm{sym}}(\rho_{B})$ above saturation density and a possible hadron-quark phase transition are among the most uncertain parts of the EoS of dense neutron~rich matter \cite{Weber07,Mark19,David20,Bao-An14}. A precise determination of the behavior of $E_{\rm{sym}}(\rho_{B})$ at densities beyond saturation is essential. To probe the  symmetry energy in this density range, data obtained from observations of NSs have revealed to be particularly useful compared with terrestrial experiments. Since the detection of GW170817, astrophysical data have stimulated many interesting studies of the symmetry energy. Recently Bao-Jun Cai et al. ~\cite{Bao-jun21} have studied the behavior of the nuclear symmetry energy at $2-3\rho_0$ on the basis of its slope $L$, its curvature $K_{\textbf{sym}}$ and its skewness $Q_{\textbf{sym}}$ at saturation $\rho_0$. This investigation was achieved by developing the function $E_{sym}(\rho_{B})$ in terms of a suitable auxiliary function, allowing an accurate prediction at high densities of the symmetry energy. 

Several attempts have been made to obtain the EOS of supernuclear matter in NSs from different models. In particular, by adopting density-dependent DD models such as DDRMF, DD-MEX~\citep{taninah2020}, DD-LZ1~\citep{wei2020}, massive NSs with masses around 2.33 - 2.48 M$_\odot$ could be generated. The DD models do not include mesonic nonlinear mixing or self-interaction terms.  Instead, a density dependence of the nucleon-meson couplings is introduced, which takes into account the effects of the nuclear medium on the couplings, as predicted, for example, by relativistic Dirac-Brueckner-Hartree-Fock calculations~\citep{brockmann1992}.  
However, DD models are generally characterized by an isovector coupling $\Gamma_\rho$ which tends to zero at higher densities. This has the consequence of softening the symmetry energy at high densities, allowing large neutron densities, and therefore these models do not predict nucleonic neutrino emission processes inside the NS that are responsible for the fast cooling of the NS, i.e. the nucleon direct Urca (DU) processes \cite{Fortin2016,Fortin:2021umb}. In order to induce the DU processes in 1.6 to 1.8 $M_\odot$ stars, as predicted in \cite{Beznogov:2015ewa}, the authors in \cite{Providencia:2018ywl,Fortin:2021umb} have included hyperons in the star. This results in a softer EoS, which makes predicting stars with masses around or above 2$M_\odot$ difficult.

Bayesian inference studies have supported the possible existence of NSs with $2.50$-$2.60\mathrm{M}_\odot$ under the constraints of NS properties of $1.4\mathrm{M}_\odot$~\citep{lim2020}.
This methodology based on microscopic models offers as a major advantage, the possibility, once the inference is complete, of discussing the  composition of nuclear matter properties through the EOS and the symmetry energy at high densities \cite{Imam2021,Mondal2021,Malik:2022zol}. It also allows the comparison of the statistical significance of a given EOS with respect to some reference EOS through the Bayes factor \cite{jeffreys1998theory, Morey_2016}, in particular to determine whether the EOS is substantially, strongly or decisively favored.  This method has recently been used in \cite{Pacilio:2021jmq} to distinguish EOSs based on different microscopic descriptions and with different compositions. Also, in \cite{Biswas2022}, the same approach was considered to determine the evidence for a set of 31 EOS using multi-messenger observations. 

The knowledge of the symmetric nuclear matter EOS has proved to be indispensable for studying the $\beta$ equilibrium condition \cite{Hebeler2013,Essick2021,Tovar2021,Mondal2021}. In this context, the authors of \cite{Malik:2022zol} have recently developed a Bayesian inference approach, in the framework of a density dependent model, in order to determine how the GW and NICER data constrain the high-density symmetry energy values. They have confirmed that these type of models do not allow the nucleon DU processes inside NSs. This is a characteristic of all density dependent models that describe  the nucleon-$\rho$ meson coupling  as an exponential decreasing function with density. However, NS cooling curves seem to indicate that these kind of processes should be possible inside NSs with a mass $\gtrsim 1.6\, M_\odot$ \cite{Beznogov:2015ewa}. 
In order to adapt these DD models at high densities, modifications have recently been introduced by Malik et al. (described in Ref.~\cite {Malik2022}) which predict nucleon DU processes inside NS. The proposed modified DD models allow to control the slope $L$  over a large range of densities both at high and low densities.  For a different parametrization of the $\rho$-meson coupling see also \cite{Scurto:2024ekq}.

In the following we propose a set of DD models, that satisfy the set of constraints on the symmetry energy proposed in several studies  \cite{Li:2021thg,Malik2022}, as well as constraints from the neutron matter calculation within a chEFT approach \cite{Hebeler2013}, which predict the onset of DU cooling processes inside the NS. In the following, we will refer as $M_{DU}$ the minimum NS mass where nucleon DU processes are allowed to occur. We will also discuss how some constraints on the properties of symmetry energy, as delineated in \cite{Li:2021thg,Oertel2017,Malik2022}, are satisfied. In particular, considering the analysis of recent NS observations, the authors of \cite{Li:2021thg} have obtained at saturation density for the symmetry energy slope $L\approx 57. 7 \pm 19$ MeV, the curvature $K_{sym}\approx -107 \pm 88$ MeV, and at twice the saturation density the symmetry energy $E_{sym}(2\rho_0) \approx 51 \pm 13$ MeV, all determined at a confidence level of $68\%$. These values are consistent with previous values obtained from other analyses based on experimental data, ab-initio calculations and NS observations, such as \cite{Li:2013ola,Lattimer:2012xj,Oertel2017}. We also consider the constraints found in \cite{Malik2022}, based on the strong correlation between the symmetry energy slope at $\rho=2.5\rho_0$ and the DU mass $M_{DU}$.  From this correlation it was found that the symmetry energy slope range at $2.5\rho_{0}$ should be in the interval $L(2.5\rho_{0})\approx 54-48$ MeV to predict DU processes occurring inside stars with a mass $M_{DU}\gtrsim 1.6-1.8\, M_\odot$.

The present study is organized as follows. In Sec.~\ref{modd} we review the formalism used in the analysis. In Sec.~\ref{disc} we discuss the properties of a set of DD models, based on the DD2 \cite{typel2010} and DDMEX \cite{Huang:2020cab} models, which predict the occurrence of nucleon DU processes within NS by modifying the parametrization of the isovector channel, keeping the same isoscalar description and the symmetry energy at saturation. In Secs.~\ref{DD} and~\ref{sec2d}, we generalize our study and apply a Bayesian inference approach to study the complete isovector channel, also modifying the symmetry energy at saturation and identifying possible correlations. Finally, some conclusions are drawn in Sec.~\ref{concl}.

\section{Models \label{modd}}

In this section, we review the description of nuclear matter EoS with relativistic mean-field (RMF) models with density dependent couplings, and the characterization of the density dependence of the symmetry energy through its expansion at the saturation density.

\subsection{Field Theoretical Models with DDRMF Lagrangian}

 In order to study the symmetry energy at high density, we will consider RMF models with density dependent baryon-meson couplings that avoid self-interacting and mixed terms between mesons and which we designate by DD models. Within this approach, we start from a Lorentz-covariant Lagrangian density which describes baryons interacting with mesons. It is assumed the minimal coupling between the baryons and the mesons. 

In the DD model, the nucleons interact with each other  by exchanging  scalar-isoscalar ($\sigma$), vector-isoscalar ($\omega^\mu$), and  vector-isovector ($\vec{\rho}^\mu$) mesons. The DDRH Lagrangian density can be written as:
\begin{align}
\mathcal{L}_{DD}=&\sum_{i=p,~n}\overline{\psi}_i\left[\gamma^{\mu}\left(i\partial_{\mu}-\Gamma_{\omega}(\rho_B)\omega_{\mu}-\frac{\Gamma_{\rho}(\rho_B)}{2}\vec{\tau}\cdot\vec{\rho}_{\mu}\right)\right. \nonumber  \\& \left.-(M-\Gamma_{\sigma}(\rho_B)\sigma) \right]\psi_i+\frac{1}{2}\left(\partial^{\mu}\sigma\partial_{\mu}\sigma-m_{\sigma}^2\sigma^2\right) \nonumber \\
&+\frac{1}{2}m_{\omega}^2\omega_{\mu}\omega^{\mu}-\frac{1}{4}\Omega^{\mu\nu}\Omega_{\mu\nu}\nonumber \\&+\frac{1}{2}m_{\rho}^2\vec{\rho}_{\mu}\cdot\vec{\rho}^{\mu}-\frac{1}{4}\mathbf B_{\mu\nu}\cdot\mathbf B^{\mu\nu},
\end{align}
where $\psi_i$ represents the Dirac spinor of the nucleons, $M$ is the nucleon mass. $\Omega_{\mu\nu}$ and $B_{\mu\nu}$ are the vector meson field tensors, $\gamma^\mu$ and $\vec{\tau}$ are the Dirac and the Pauli matrices, respectively. The $\Gamma_\sigma(\rho_B)$, $\Gamma_\omega(\rho_B)$ and $\Gamma_\rho(\rho_B)$ are the coupling constants of the nucleons to the meson fields $\sigma$, $\omega$, and $\rho$ respectively, with a corresponding meson masses are $m_\sigma,\, m_\omega$ and $m_\rho$.

We focus on the parametrizations of DDRMF  that depend on the vector density.
The nucleon-meson density-dependent coupling parameters are written in the form: 
\begin{equation}
\Gamma_{j}(\rho_B) = \Gamma_{j}(\rho_0) h_M(x)~,\quad x = \rho_B/\rho_{0}~,\quad j =\sigma ,\  \omega,\ \rho 
\label{coup}
\end{equation}
where $\rho_0$ is the saturation density of
symmetric nuclear matter. In the present study, we consider the 
parametrizations DD2 \cite{typel2020} and DDMEX \cite{taninah2020}. 
For these two parametrizations, the coupling constants of the $\sigma$ and $\omega$ mesons to the nucleons are  given in terms of the functions $h_j$ ~\cite{typel2020},
\begin{equation}
h_j(x) = a_j \frac{1 + b_j ( x + d_j)^2}{1 + c_j (x + d_j)^2},
\end{equation}
with $x=\rho_B/\rho_0$ and $\rho_B$ the baryonic  density. The parameters $a_j, b_j, c_j,$ and $d_j$ are defined in 
 \cite{typel2020} for DD2 and  \cite{taninah2020} for DDMEX. 
In the DD2 and DDMEX models, the nucleon DU processes do not occur in the NS interior because the $\rho_{DU}$ density is greater than the central density of the most massive star, $\rho_c$. In order to overcome this property  in \cite{Malik2022} a generalization of the isovector channel $\rho$-meson coupling was proposed, including a new parameter $y$ which controls the high-density behavior. This parametrization will be considered in the present study:
\begin{equation}
h_\rho(x) = y ~ \exp[-a_\rho (x-1)] + (1-y) ~, \quad 0<y\le 1~.
\label{hm2}
\end{equation}

From the Lagrangian density, we obtain the following meson field equations in the mean-field approximation
\begin{align}
\label{1.8}
&m_{\sigma}^2\sigma=\Gamma_{\sigma}(\rho_B)\rho_s=\Gamma_{\sigma}(\rho_B)\left\langle\overline{\psi}\psi\right\rangle,\nonumber\\
&m_{\omega}^2\omega=\Gamma_{\omega}(\rho_B)\rho_B=\Gamma_{\omega}(\rho_B)\left\langle\psi^{\dag}\psi\right\rangle,\nonumber\\
&m_{\rho}^2\rho=\frac{\Gamma_{\rho}(\rho_B)}{2}\rho_3=\frac{\Gamma_{\rho}(\rho_B)}{2}\left\langle\psi^{\dag}\tau_3\psi\right\rangle,
\end{align} 
where $\rho_s$ is the scalar density, and $\rho_3=\rho_p-\rho_n$ with $\rho_p$ and $\rho_n$, respectively, the proton and the neutron densities.
The Dirac equations for the nucleons are given by 
\begin{align}
\label{1.9}
 &\left[i\gamma^{\mu}\partial_{\mu}-\gamma^0\left(\Gamma_{\omega}(\rho_B)\omega+\frac{\Gamma_{\rho}(\rho_B)}{2}\rho\tau_3+\Sigma^R_0(\rho_B)\right)\right. \nonumber\\ &\left.-M_B^{*}\right]\psi_B=0,
\end{align} 
where $\omega$ and $\rho$ define the time components of $\omega^\mu$ and of the third component of $\vec \rho^\mu$, 
 the nucleon isospin third components  take the values $\tau_3=1$ and $\tau_3=-1$ for protons and neutrons, respectively,  the effective mass of nucleons is given in terms of the scalar meson $\sigma$ 
\begin{align}
M_B^{*}&=M-\Gamma_{\sigma}(\rho_B)\sigma.
\end{align}
and the rearrangement term $\Sigma_0^R$, due to the density dependence of coupling constants \cite{typel2020, Lalazissis2005, Antic2015},  is given by
\begin{gather}
\Sigma^R_0(\rho_B)=\frac{\partial\Gamma_{\omega}(\rho_B)}{\partial\rho_B}\omega\rho_B+\frac{1}{2}\frac{\partial\Gamma_{\rho}(\rho_B)}{\partial\rho_B}\rho\rho_3-\frac{\partial\Gamma_{\sigma}(\rho_B)}{\partial\rho_B}\sigma\rho_s.
\end{gather}

From the energy-momentum tensor in a uniform system, the energy density, $\mathcal{E}$, and the pressure, $P$, of infinite nuclear matter can be obtained, respectively, as
\be
\mathcal{E}&=&\frac{1}{2}m_{\sigma}^2\sigma^2+\frac{1}{2}m_{\omega}^2\omega^2+\frac{1}{2}m_{\rho}^2\rho^2 \cr
&+& \sum _{i=p,n}\frac{\gamma}{2\pi^2}\int_{0}^{k_{Fi}}k^2\sqrt{k^2+{M_i^{*}}^{2}}dk,
\ee
\be
P&=&\rho_B\Sigma_{R}(\rho_B)-\frac{1}{2}m_{\sigma}^2\sigma^2+\frac{1}{2}m_{\omega}^2\omega^2+\frac{1}{2}m_{\rho}^2\rho^2\cr
       &+&\sum _{i=p,n}\frac{\gamma}{6\pi^2}\int_{0}^{k_{Fi}}\frac{k^4 dk}{\sqrt{k^2+{M_i^{*}}^{2}}},
\ee
where  $\gamma=2$ is the spin degeneracy factor, and $k_{Fi}$ is the Fermi momentum of nucleon $i$. The binding energy per nucleon is defined by
\begin{equation}
\frac{E_{b}}{A}=\frac{\mathcal{E}}{\rho_{B}}-M.
\end{equation}
\subsection{Nuclear symmetry energy}
For  nuclear matter consisting of protons and neutrons at the density $\rho_B$ with the isospin asymmetry parameter $t=(\rho_{n}-\rho_{p})/\rho_B$, the energy per baryon(nucleon) $E=\displaystyle\frac{\mathcal{E}}{\rho_{B}}$, can be expanded with respect to $t$ and $\rho_B$:
\begin{equation}
{E}(\rho_{B},\delta)\simeq E_{SNM}(\rho_{B})+E_{sym}(\rho_{B})t^2+ {\cal O}(t^{4})  \ ,
\label{ea}
\end{equation}
where $E_{SNM}(\rho_{B})$ is the energy per particle of symmetric nuclear matter and  the symmetry energy $ E_{sym}(\rho_{B})$ quantifies the energy needed to make nuclear matter more neutron rich. Assuming the charge symmetry of the nuclear forces 
the symmetry energy can be identified as a quadratic term in $t$ of the energy per particle: 
\begin{equation}
E_{sym}(\rho_{B}) =\frac{1}{2}\frac{\partial^2E}{\partial t^2}\Big
|_{t=0}.
\label{ea2}
\end{equation}
To study the density dependence in a range of densities below the quark-hadron phase transition, it is common to characterize the density dependence of the~SNM EOS $E_{0}(\rho_{B})$ and the symmetry energy $E_{\rm{sym}}(\rho_{B})$ around the saturation density $\rho_0$ in terms of the parameters of the Taylor series. In the following, we consider terms until third order on the variable $\displaystyle x=\frac{\rho_{B}-\rho_0}{3\rho_0}$, 
\begin{eqnarray}
E_{0}(\rho_{B})&=&E_0(\rho_0)+\frac{K_0}{2}x^2+\frac{Q_0}{6}x^3,\\
E_{\rm{sym}}(\rho_{B})&=&E_{\rm{sym}}(\rho_0)+L_{0}\, x+\frac{K_{\rm{sym},0}}{2}\,x^2 \cr
&+&\frac{Q_{\rm{sym},0}}{6}x^3,
\label{ea3}
\end{eqnarray}
$E_0(\rho_{B})\equiv E_{\rm{SNM}}(\rho_{B})$ is the  energy per nucleon for symmetric nuclear matter (SNM). The incompreensibility  coefficient $K_0$ and the the skewness $Q_0$ are, respectively, defined by: 
\begin{eqnarray}
K_0&=&9\rho_0^2\dfrac{\partial^2 E}{\partial 
\rho_{B}^2} \Bigg|_{\rho_{B}=\rho_0}=9\left[\frac{dP_{0}}{d\rho_B}\right]_{\rho_B=\rho_0}
\nonumber \\
Q_0&=&27\rho_0^3\dfrac{\partial^3 E}{\partial 
\rho_{B}^3} \Bigg|_{\rho_{B}=\rho_0}\nonumber\\
\label{ceng}
\end{eqnarray}
with $P_{0}$ the SNM pressure .
$E_{\rm{sym}}(\rho_0)$ is  the symmetry energy at saturation and the
quantities $L_0$, $K_{sym}$ and $Q_{sym}$  are  its slope, curvature and skewness respectively, at saturation:
\begin{equation}
\begin{array}{c}
\displaystyle{L_{0}=3\rho_0\frac{\partial E_{sym}(\rho_{B})}{\partial \rho_{B}} \Big |_{\rho_{B}=\rho_0}} \ , \nonumber\\
\displaystyle{K_{sym,0}=9\rho_0^2\frac{\partial^2 E_{sym}(\rho_{B})}{\partial \rho_{B}^2} \Big |_{\rho_{B}=\rho_0}}\, \nonumber\\ \    \ \displaystyle{Q_{sym,0}=27\rho_0^3\frac{\partial^3 E_{sym}(\rho_{B})}{\partial \rho_{B}^3} \Big |_{\rho_{B}=\rho_0}}.
\end{array}
\label{ea4}
\end{equation}
This is  considered to be a good approximation even for small proton 
fraction $y_p$ \cite{mishra2015}. 

These parameters characterize the density dependence of nuclear symmetry 
energy around normal nuclear matter density and thus provide important 
information on the behavior of nuclear symmetry energy at both high and 
low densities. Also, the curvature parameter $K_{sym,0}$ would be a significant measurement which distinguishes the different parametrizations. 
The shift of the incompressibility with asymmetry, is given by
\begin{equation}
K_{asy,0}=K_{sym,0}-6L_{0} -\frac{Q_0}{K_0}L_{0}\approx K_{sym,0}-6L_{0}.
\end{equation}
This value can be correlated to experimental observations of the  giant monopole resonance (GMR) of neutron-rich nuclei \cite{tili} such as  on even-$A$ Sn isotopes with a value of  $K_{asy,0}=-550 \pm 100$ MeV, see discussion \cite{Chen:2004si,Li:2008gp,Centelles:2008vu}.

\section{Results and discussion \label{disc}}

The parameterisation of the $\rho$ meson coupling proposed in Eq. (\ref{hm2}) allows us to obtain a set of modified DD models consistent with both low and high density constraints. We obtain for a given symmetry energy at saturation density  a range of different allowed isovector properties, in particular the symmetry energy slope $L_0$ and its curvature $K_{sym,0}$, for the same  isoscalar properties i.e incompressibility, binding energy and effective mass of the baryons as shown in Table~\ref{Tab1a}.

The proton fraction and other NS characteristics which are sensitive to the high density behavior of the symmetry energy are used to constrain the new parameter, $y$, in Eq. (\ref{hm2}). In particular, the onset of the nucleon DU processes as discussed in \cite{Beznogov:2015ewa} is an observation that  constrains $y$, see \cite{Malik2022}.

\begin{figure*}[h]
\begin{center}
\hspace*{-1.0cm}
\begin{tabular}{cc}
\includegraphics[width=0.54\linewidth]{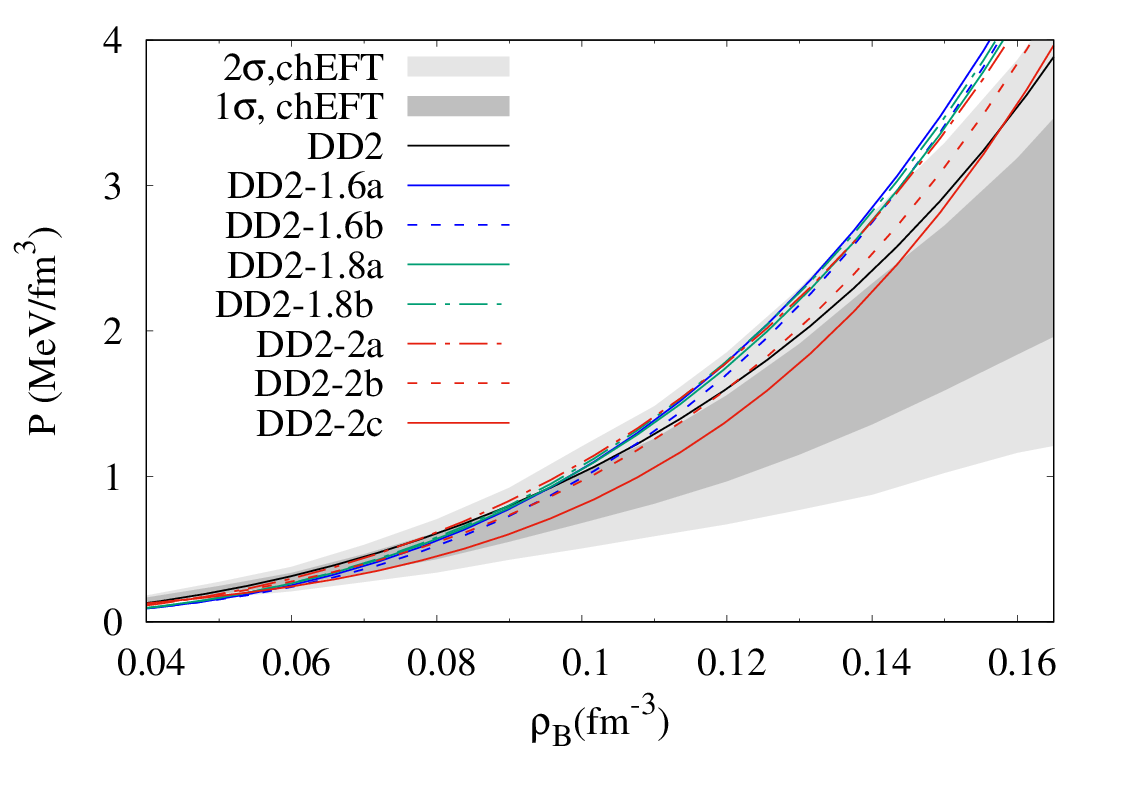}&
\includegraphics[width=0.54\linewidth]{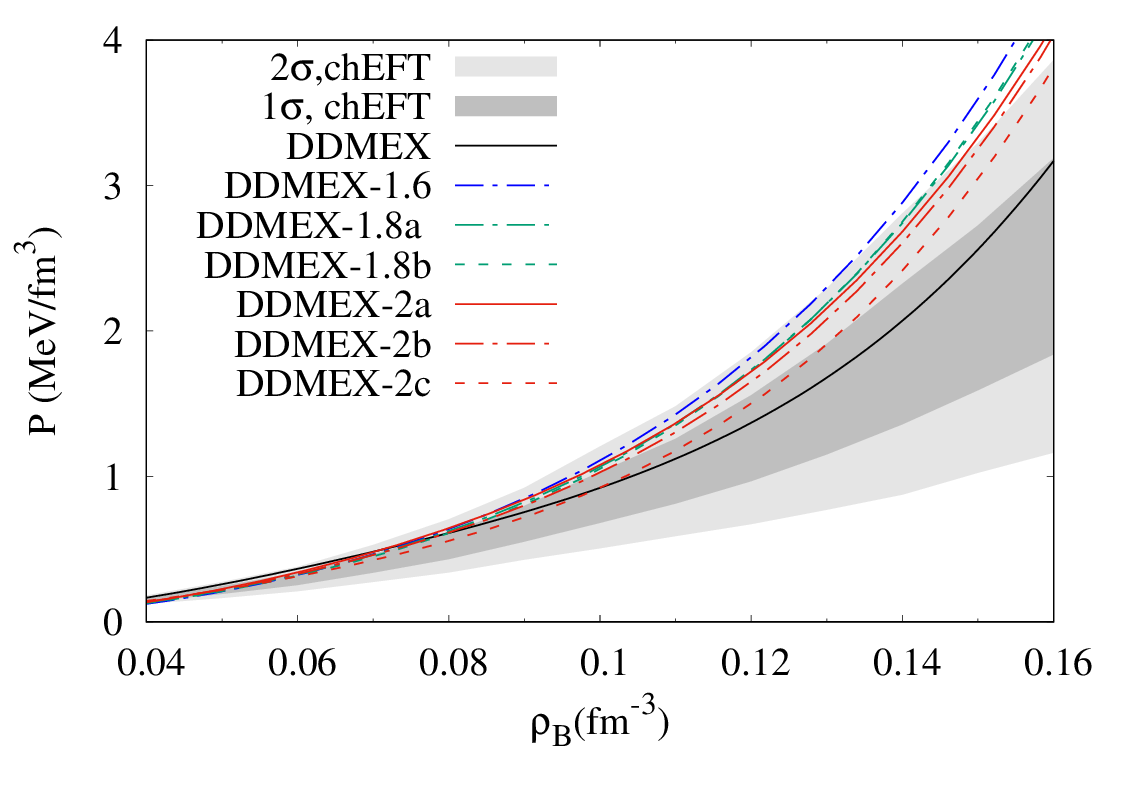}\\
\includegraphics[width=0.50\linewidth]{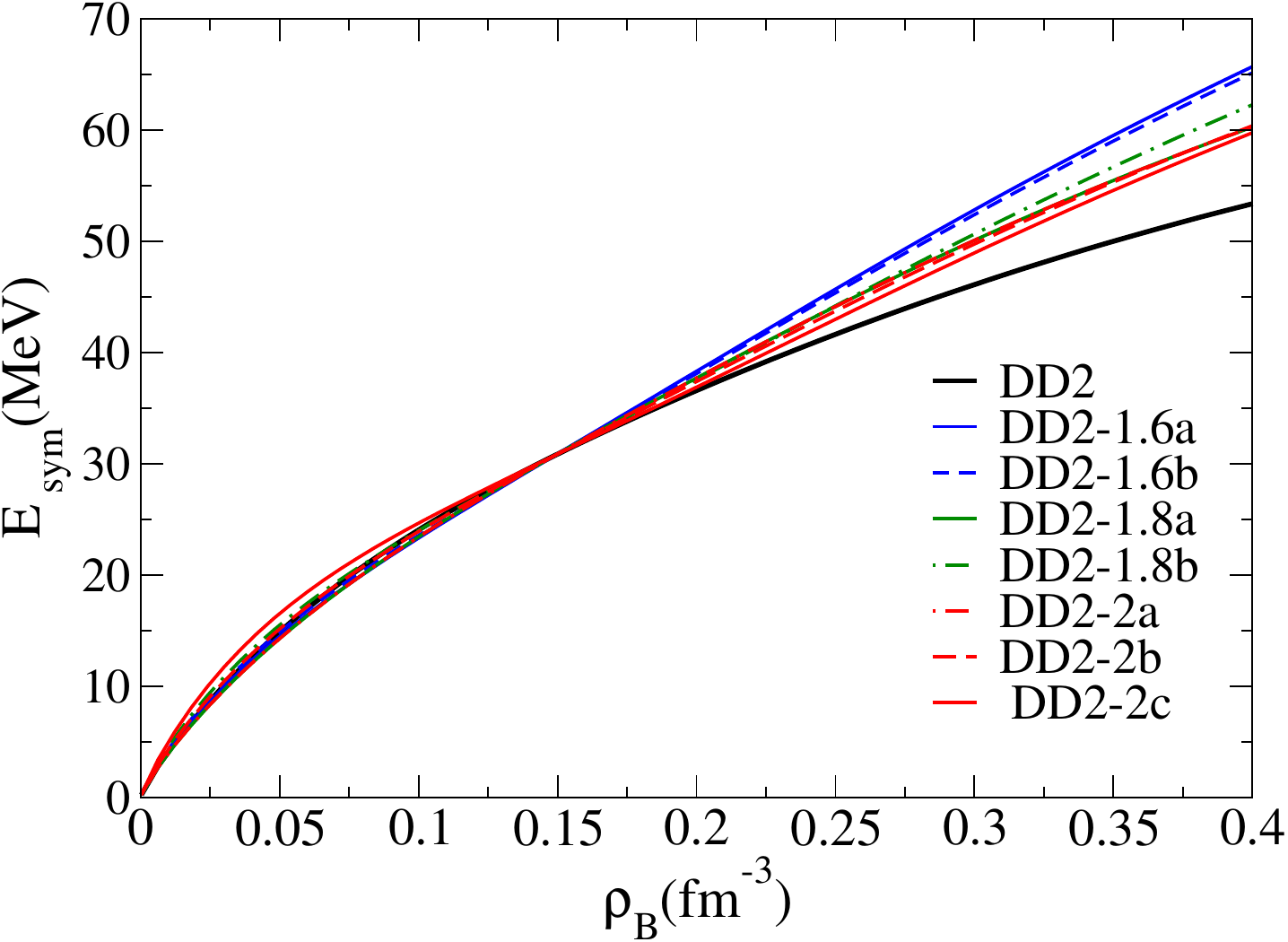}&
\includegraphics[width=0.50\linewidth]{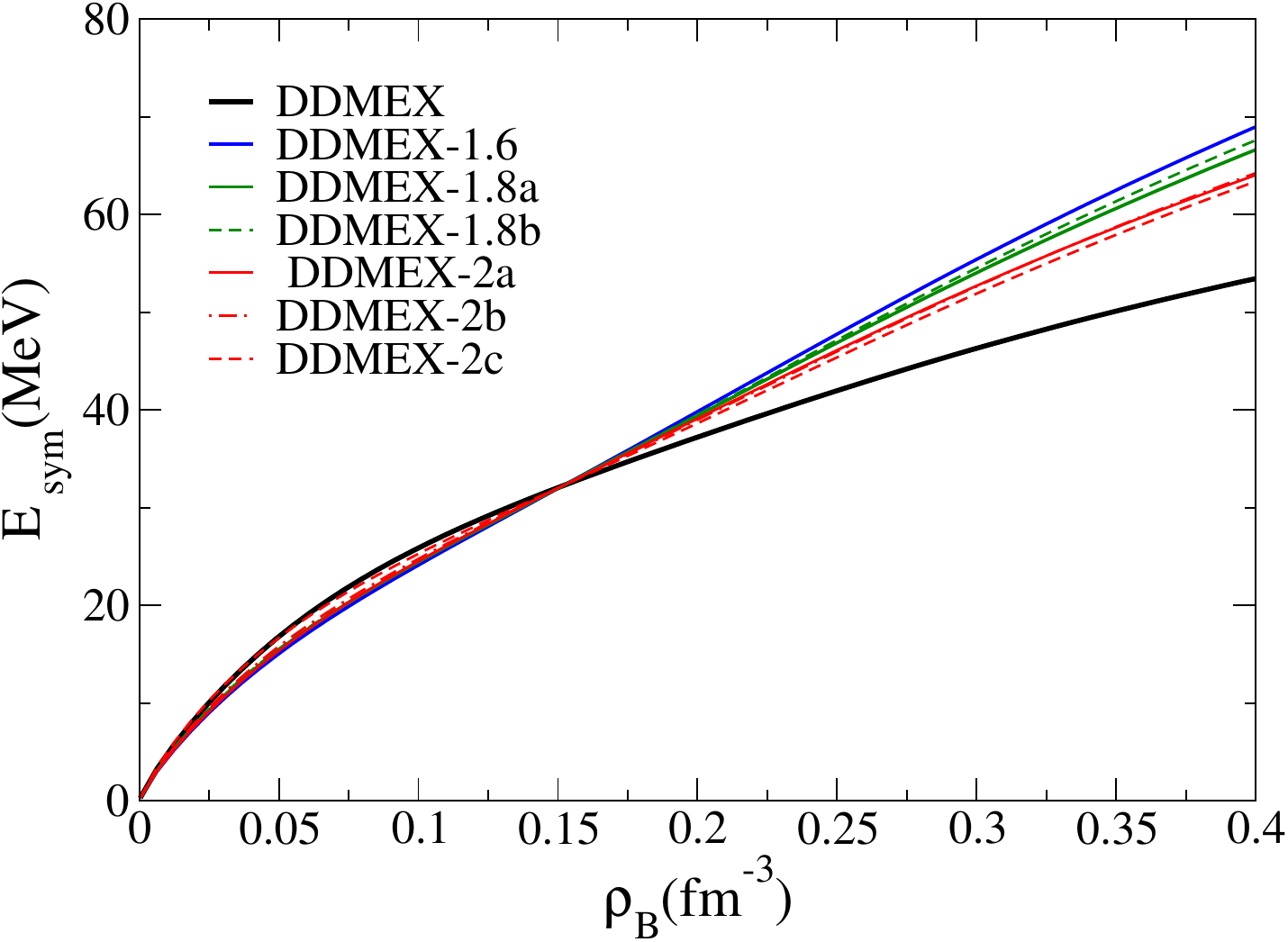}\\
\includegraphics[width=0.50\linewidth]{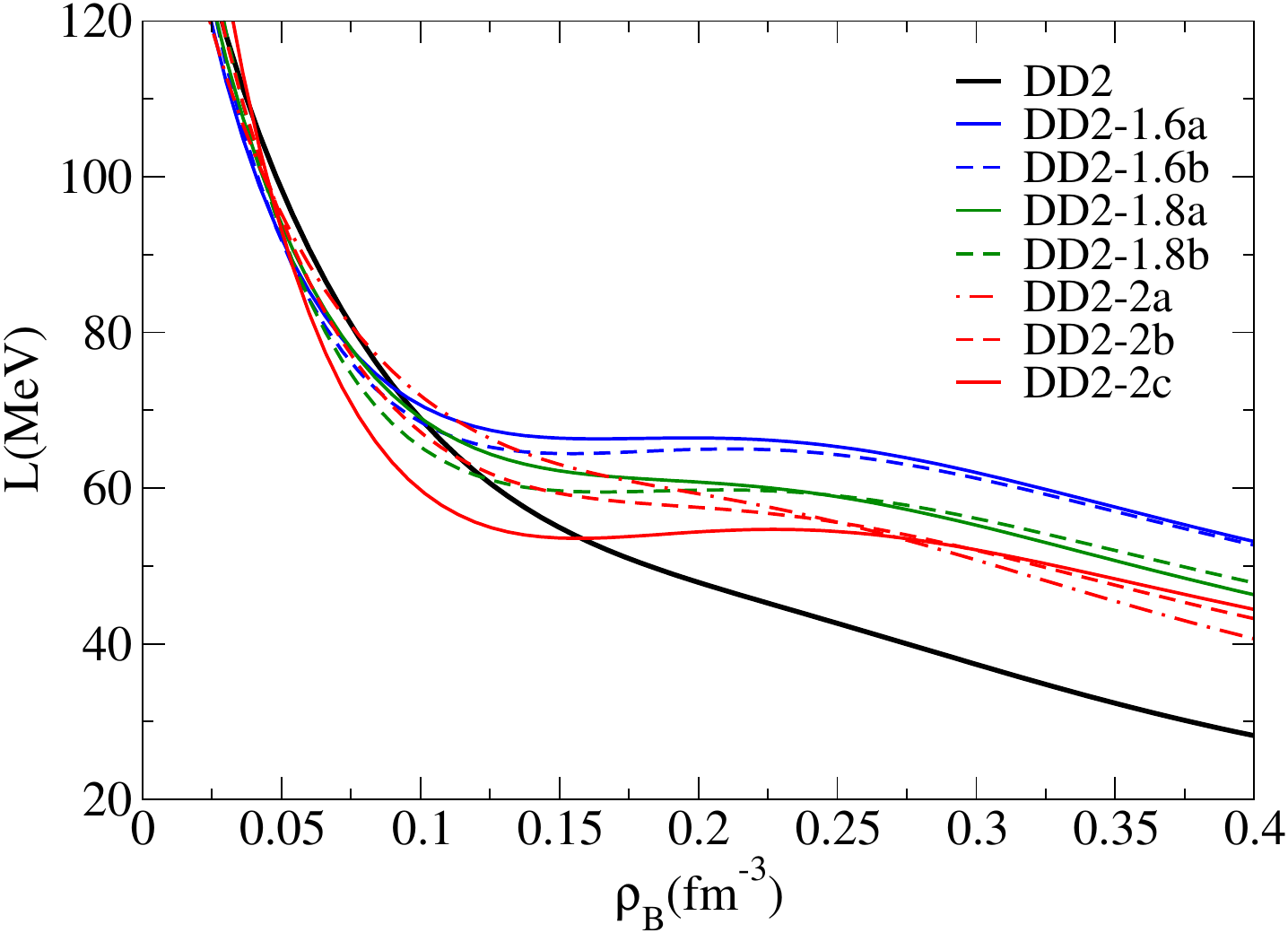}&
\includegraphics[width=0.50\linewidth]{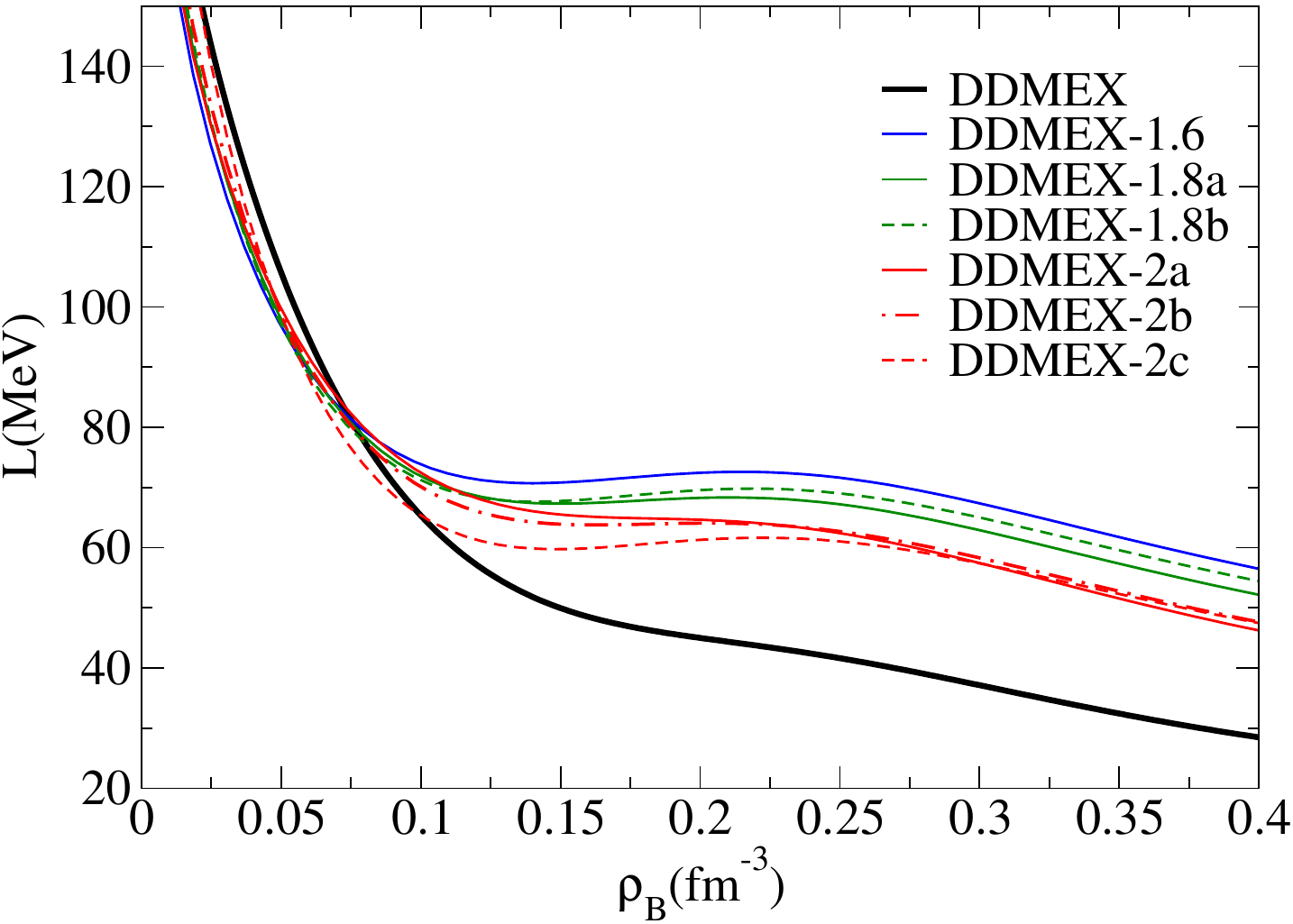}
\end{tabular} 
\caption{(Color online) Upper panels show the pure neutron matter pressure as a function of the baryonic density for DD2 (left) and DDMEX (right) families under consideration varying ($a_{\rho}, y$) to obtain a given M$_{DU}$ i.e. 1.6, 1.8, 2.0 M$_\odot$ and including the 1$\sigma$ and 2$\sigma$ bands from chiral effective field theoretical calculations \cite{Hebeler2013}. Middle panels reveal the symmetry energy and in the bottom its slope as function of the baryon density for DD2 (left) and DDMEX (right) families.}
\label{fig13}
\end{center}
\end{figure*}

We build a set of EOS based on the  modified DD2 and DDMEX models that essentially satisfy neutron matter chiral effective field theory (chEFT)~\cite{Hebeler2013,Drischler:2015eba} constraints within $2\sigma$, as shown in  Fig.~\ref{fig13} top panels, and verify the constraints on the symmetry energy at saturation density referred in the Introduction  \cite{Li:2021thg,Li:2013ola,Lattimer:2012xj,Malik2022}: {a slope of nuclear symmetry energy at saturation density $\rho_0$  satisfying $L_0\approx 57.7\pm 19$ MeV at 68\% confidence level and a symmetry energy at twice $\rho_0$, $E_{\rm{sym}}(2\rho_0)\approx 51\pm 13$ MeV at 68\% confidence level \cite{Li:2021thg};
the symmetry energy and respective slope satisfying  $E_{sym} (\rho_0) = (31.7 \pm 3.2)$~MeV and $L_{0} = (58.7 \pm 28.1)$~MeV \cite{Oertel2017}; the  slope  $L$ for a density $\approx 2.5 \rho_{0} $  taking  a value in the range  $48 - 54$ MeV for the nucleon DU processes to occur inside neutron stars \cite{Malik2022}.}

ChEFT constraints are generally considered  when modeling the EOS of neutron stars, specially below saturation density.
Note, however, that for densities  of the order of $\rho_0$ and above  the uncertainties associated with the chEFT neutron matter become quite  large because the low-momentum expansion breaks down. In addition, different treatments of the three-body force result in different uncertainties associated with the different calculations, see \cite{Hebeler2013,Lynn:2015jua,Drischler:2017wtt,Raaijmakers_2021,Huth:2021bsp}. Therefore, we did not take the chEFT constraint strictly, especially near $\rho_0$.  To test the EOS proposed in this subsection, a Bayesian inference is performed and the Bayes factor of these EOS is determined with respect to the EOS with the highest likelihood within each inference model, see subsections \ref{DD} and \ref{sec2d}. 

In order to constraint the high density behavior of the symmetry energy, the parameters  $(y,a_{\rho})$ were chosen such that a given $M_{DU}$ is obtained, i.e. 1.6, 1.8, 2.0 $M_\odot$. 
Acceptable models should also satisfy the low density constraints imposed by chEFT neutron matter calculations.
In Table \ref{Tab11}, some symmetric nuclear matter properties calculated at saturation density and at $2.5\rho_0$ are given for all proposed EOS. As referred before, by construction the two families of models DD2 and DDMEX have the symmetry energy at saturation $E_{sym,0}$ of the original DD2 and DDMEX, given in Table~\ref{Tab1a}.

By analyzing Table~\ref{Tab11}, it is seen that $L_0$ is correlated (anti-correlated) with $y$ ($a_\rho$) while
$L(2.5\rho_0)$ are anti-correlated (correlated) with $y$ ($a_\rho$): 
 The combined effect of the two parameters is to {increase} the slope of $L$ at high densities, preventing the symmetry energy from softening.
On the other hand, $K_{sym,0}$ is correlated with $a_\rho$ and  anti-correlated  with $y$: it increases when $a_\rho$ increases and remains negative for DD2, while it can reach positive values for DDMEX.
Recently, positive values for $K_{sym,0}$ have been obtained in ~\cite{Reed2023} in an attempt to simultaneously describe the results of PREX2 \cite{PREX:2021umo} and CREX \cite{CREX:2022kgg} within a model that included the $\delta$-meson. %
However,  previous analysis seem to prefer negative values: in \cite{Li:2021thg}, the authors predict $K_{sym,0}\equiv -107\pm 88$~MeV at 68\% confidence level, considering a large set observational data, a value consistent with other  analyses that also considered experimental data -100$\pm 100$~MeV \cite{Margueron:2017eqc} and -112$\pm 71$~MeV \cite{Mondal:2017hnh}.  

{In general, the effect of including the parameter  $y$ is to increase the symmetry energy slope $L_0$ at saturation. However, an adequate  choice of both parameters $(y,\, a_\rho)$ also allows to fix $L_0$ with a value smaller than the one of DD2, but with a much larger value of $K_{sym,0}$. For the DDMEX family we have only obtained EOS with  $L_{0}>49$~MeV, the value that characterizes DDMEX.}

In the Fig. \ref{fig13} (top panels), the pure neutron matter (PNM) chEFT  EOS from \cite{Hebeler2013} is included considering $1\sigma$ (dark gray) and $2\sigma$ (light gray), as well as the PNM EOS of the DD modified models for $M_{DU}$ equal to 1.6, 1.8 and 2.0 $M_\odot$, the DD2 family in the left panel and the DDMEX in the right one. {The constraints derived for PNM exclude DU processes inside stars with a mass $\lesssim$ 1.6 $M_\odot$ at $2\sigma$. According to \cite{Beznogov:2015ewa}, the  NS cooling curves seem to indicate that  M$_{\rm DU}$ $\sim 1.6 - 1.8$ M$_\odot$. Within the DDMEX family we are not able to go below M$_{\rm DU}=1.8 M_\odot$ and simultaneously satisfy chEFT constraints. It is seen that within the two families and the parametrization defined in Eq. (\ref{hm2}) for the $\rho$-meson coupling, the chEFT constraints of \cite{Hebeler2013} are only satisfied in the complete density range within the 2$\sigma$ interval. Introducing the $y$ contribution makes the symmetry energy harder above saturation density, as expected since this term does not allow the $\rho$-meson coupling to converge to zero, see the middle panels. The bottom panels show the symmetry energy slope as a function of the density and the effect of the $y$ parameter is easily identified giving rise to a kind of plateau before a soft decrease occurs at the larger densities.}

\begin{table}[htb]
\caption{\label{tab1}
The properties of symmetric nuclear matter at saturation density for the models under study: the nuclear saturation density $\rho_0$, the binding energy per particle $B/A$, the incompressibility $K_0$, the skewness coefficient $Q_{0}$, the symmetry energy $E_{sym,0}$, the slope of the symmetry energy $L_0$, and the effective nucleon mass $M^{*}$. All quantities are in MeV, except $\rho_0$ which is in fm$^{-3}$, and the effective nucleon mass is normalized to the nucleon mass.}\begin{ruledtabular}
\vspace{0.5cm}
\begin{tabular}{cccccccc}
 Model  &$\rho_0$ &$B/A$ &$K_0$ & $ E_{sym,0}$ & $M^{*}/M$ & $ Q_{0}$\\
\hline
DD2 Family & 0.15   & -16.03 & 243 &  30.8 & 0.55& 169\\
DDMEX Family & 0.15  & -16.09  &267  &32.3 &0.55 &  877\\
\end{tabular}
\label{Tab1a}
\end{ruledtabular}
\end{table}

\begin{table}[h]
\caption{The symmetric nuclear matter properties at saturation density for the models under study: The slope of the symmetry energy $L_0$ and the curvature $K_{sym,0}$ and $K_{asy,0}$ at saturation density,  the slope of the symmetry energy $L$ at $2.5$ saturation density  for the models under study for a set of ($a_{\rho}$,y). 
}
\begin{ruledtabular}
\vspace{0.5cm}
\begin{tabular}{lcccccc}
 Model    &$a_\rho$ &y & $L_0$ & $L(2.5 \rho_0)$& $ K_{sym,0}$ & $K_{asy,0}$\\
\cline{4-7} 
         &      &    &   \multicolumn{4}{c}{MeV}\\ 
\hline
DD2 family: &      \multicolumn{2}{c}{}\\
DD2      &0.5189  &1.0  & 55.12& 30.27& -93.21 & -415.93 \\
DD2-2a & 0.8 &0.51 & 63.85& 43.16& -52.18& -429.84 \\
DD2-2b & 1.0 &0.47 & 59.38& 45.53& -34.93& -386.78  \\
DD2-2c  & 1.2 &0.45 & 53.59& 46.60& -9.62 &  -328.576 \\
DD2-1.8a &1.05 &0.4  & 62.18& 48.60& -28.76 &  -398.01 \\
DD2-1.8b & 1.2 &0.38 & 59.55& 50.05& -13.45& -367.96  \\
DD2-1.6a  & 1.25&0.29 & 66.33& 55.45& -9.636& -405.16  \\
DD2-1.6b &1.3  &0.3  & 64.50& 54.95& -5.078 & -406.86 \\
\hline
DDMEX family:& \multicolumn{2}{c}{}\\
DDMEX     &0.6202&1.0  &49.66  &29.87 &-71.57 & -369.17\\
DDMEX-2a  &1.0&0.42 &65.41  &48.27 &-8.08 & -410 \\
DDMEX-2b  &1.1&0.4  &63.85  &49.60 &-8.09  & -390.84 \\
DDMEX-2c  &1.2&0.41 &59.75  &49.34 & 3.92  & -354.25\\
DDMEX-1.8a  &1.2&0.33 &67.31  &54.18 & 1.92  & -401.68\\
DDMEX-1.8b &1.3&0.3  &67.80  &56.38 & 11.1  & -395.4\\
DDMEX-1.6  &1.3&0.27 &70.86  &58.53 & 10.78 & -411.88 \\
\end{tabular}
\label{Tab11}
\end{ruledtabular}
\end{table}

By varying $(y, a_\rho)$ it is possible to increase the slope of the symmetry energy at $2.5\rho_0$   with respect to the original models, DD2 and DDMEX: $L(2.5\rho_0$) varies in the range [43,55]~MeV for DD2 family and [48,58]~MeV for DDMEX family. This intervals are reasonably compatible with $L(2.5\rho_0)\equiv 48-54$~MeV, taking into account that they have been deduced from the strong correlation existing between $L$ and M$_{\rm DU}$ \cite{Malik2022},
\begin{equation}
 \frac{L}{\rm MeV}=\frac{-31.224}{\rm MeV}\frac{M_{\rm DU}}{M_\odot}+\frac{104.339}{\rm MeV}
\end{equation}
with a Pearson coefficient of 0.9. 
\begin{figure}[htb]
\begin{center}
\includegraphics[angle=0, width=1.02\linewidth]{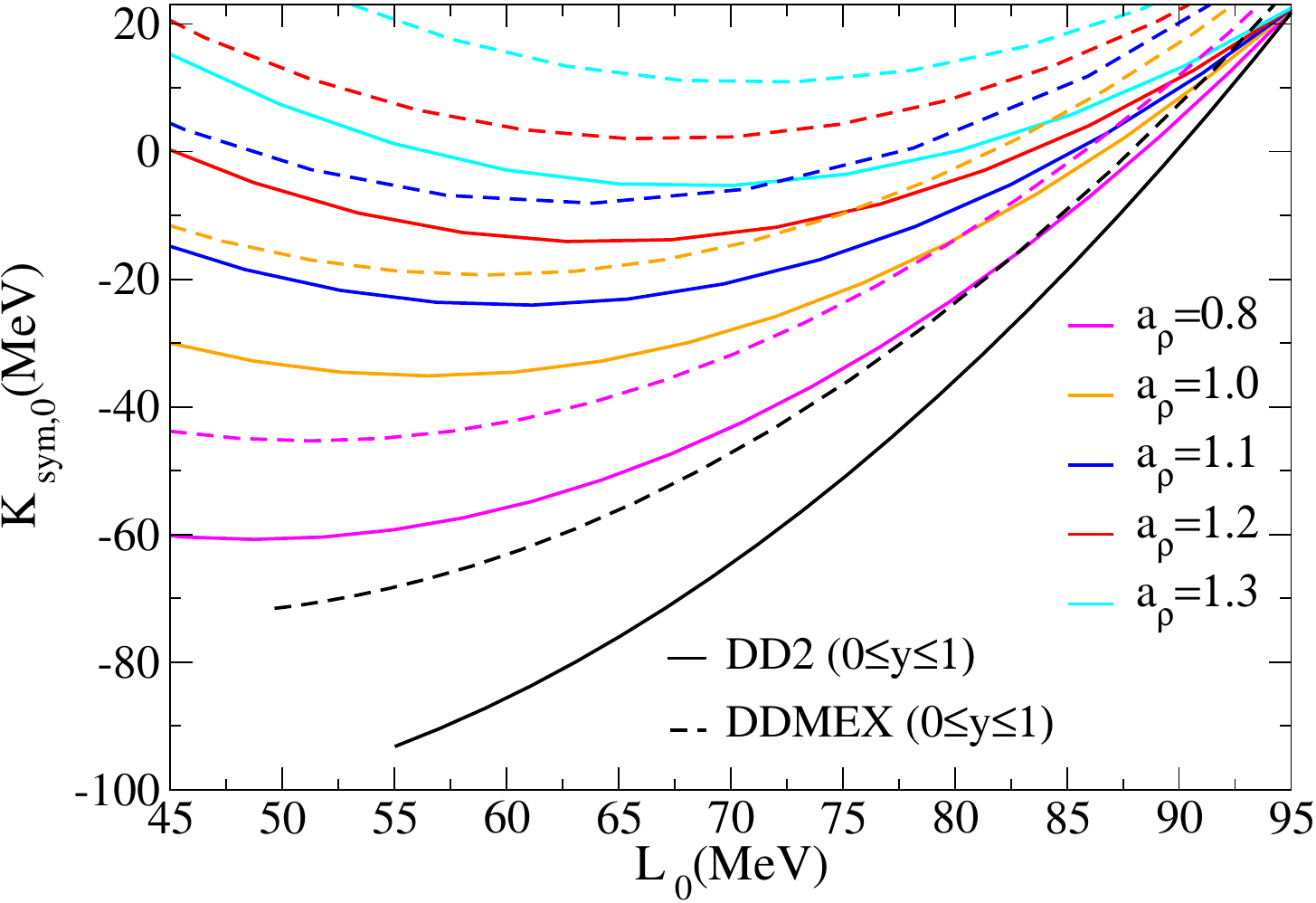}\\
\end{center}
\caption{$K_{sym,0}$ versus $L_{0}$. Original DD2 (black solid line) and DDMEX (black dashed line) varying $y$, modified DD2 (colored solid line) and DDMEX (colored dashed line) using different values of the parameter $a_\rho$ and varying $y$. %
} 
\label{ksym}
\end{figure}

 DDMEX model presents a softer EoS with a lower slope L$_{0}=49.7$MeV that satisfies the 1$\sigma$ constraint as shown in Fig.\ref{fig13}. We notice, the higher the slope the stiffer is the pressure. The DDMEX-1.6  model has  L$_{0}=70.9$MeV and misses the chiral EFT above  $0.13 $fm$^{-3}$. 

 In Fig.~\ref{ksym}, $K_{\rm sym,0}$ is plotted  versus $L_0$  changing the parameter $y$ between 0 and 1, and considering several values for the $a_\rho$ parameter.  A negative correlation between $L_0$ and $K_{\rm  sym,0}$ is obtained when $a_\rho$ takes the original values of models  DD2\cite{typel2010} and DDMEX \cite{Huang:2020cab}.  A similar correlation was obtained in \cite{Providencia14,Ducoin2010,Ducoin2011}, where it was shown that the coefficients $L_0$ and $K_{\rm sym,0}$  taken at saturation density for a large set of RMF and Skyrme force models  present a linear  anti-correlation between them.
 However, for the modified EOS with a larger parameter $a_\rho$,  the $L_0$-$K_{\rm  sym,0}$ variation is not monotonic, showing that the $L_0$-$K_{\rm  sym,0}$ correlation is washed out for sufficiently large values of the parameter $a_{\rho}$.

\subsection{Direct Urca process with modified DD models}
In this section, we investigate the opening density of the direct DU process for the modified density-dependent model families. The nucleonic DU processes \cite{Lattimer:1991ib} by neutrino emission are described by the equations :
\begin{equation}
n \rightarrow p + e^- + \bar{\nu}_e, \qquad \textrm{and} \qquad p+e^- \rightarrow n+\nu_e.
\label{eqn:DUrcanpe}
\end{equation}
This process is considered to be the most efficient NS cooling mechanism. It is only triggered if there is conservation of momentum, i.e. $p_{{\rm F}n}\leq p_{{\rm F}p} + p_{{\rm F}e}$, where $p_{\rm Fi}$ is the Fermi momentum of the particle species $i$. 
Therefore, the proton fraction needs to equal or exceed a minimum proton fraction $y_p^{\rm min}$ \cite{Klahn06}, to allow the DU process to operate:
\begin{equation}
y_{p}^{\rm min}=\frac{1}{1+\left(1+x_e^{1/3}\right)^{3}},
\end{equation}
with $x_e=\rho_e/\left(\rho_e+\rho_\mu\right)$, and $\rho_e$ and $\rho_\mu$ the electron and muon densities. 
In the following, $\rho_{\rm DU}$  denotes the baryon density at which the DU process starts to operate  and $M_{\rm DU}$ is the mass of the star with a central density equal to $\rho_{\rm DU}$.

\begin{figure}[h]
\begin{center}
\includegraphics[width=1.04\linewidth]{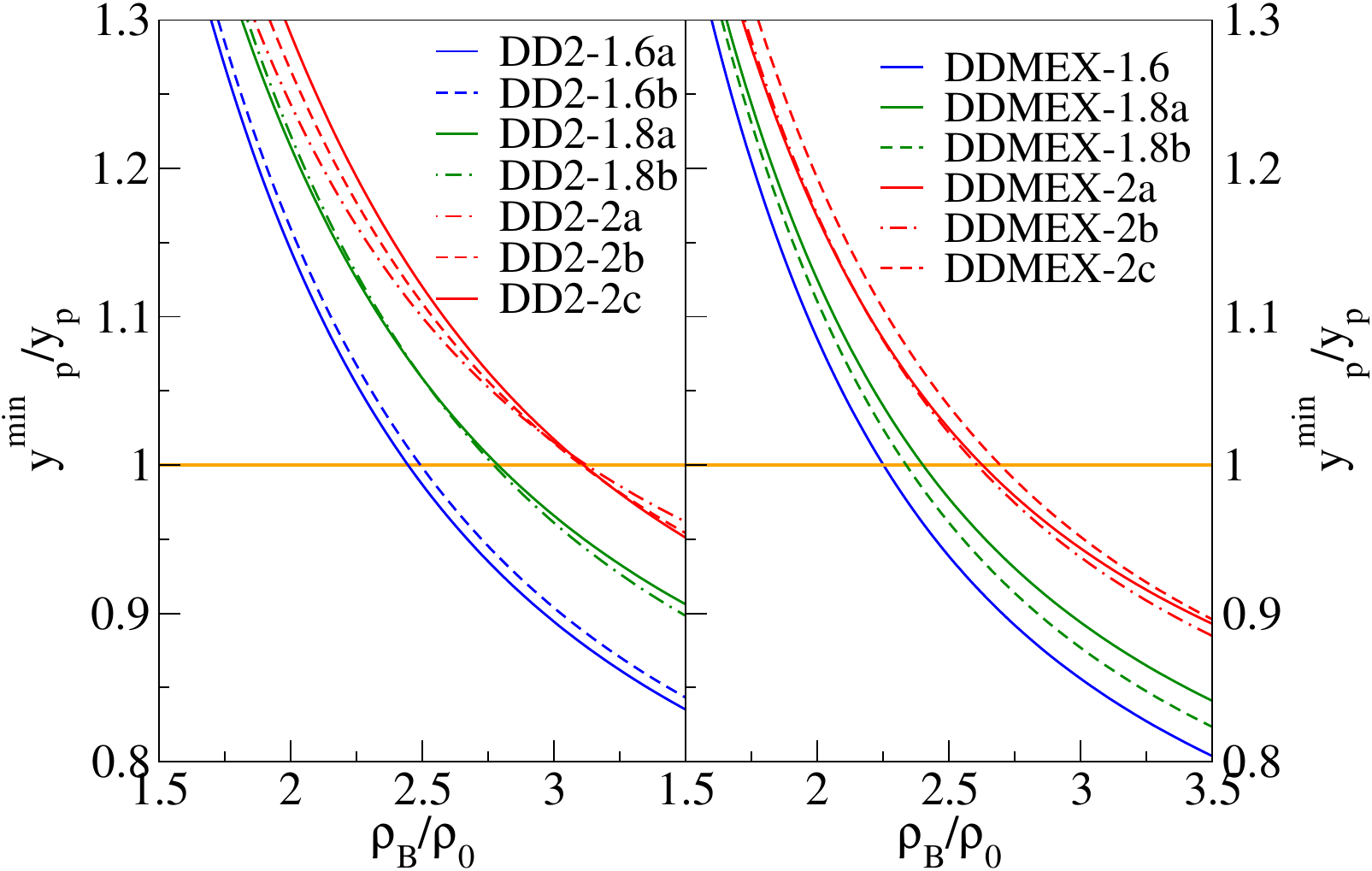}
\caption{(Color online) The DU onset for a given M$_{DU}$, i.e 1.6, 1.8, 2.0 $M_\odot$ (blue, green, red), for the DD2 family (left) and the  DDMEX family (right).}
\label{fig12}
\end{center}
\end{figure}

In previous studies~\cite{Cavagnoli11,Providencia14}, the authors investigated how the symmetry energy in various non-linear RMF models affects the DU process, emphasizing its connection to the density dependence of symmetry energy. They discovered that this dependence notably influences this process. In cases like DD2 and DDMEX models, nucleonic DU processes are prohibited, as in most conventional DD models. However, fast cooling can still occur with hyperon onset. In such cases, hyperonic DU processes, involving hyperon decay into other hyperons or nucleons with neutrino emission, begin just above hyperon onset densities \cite{Fortin2016}.

To explore the impact of the density dependence of the symmetry energy slope $L$ on the DU processes, we depict in Fig.~\ref{fig12} the onset density of nuclear DU processes 
the DD modified models discussed in Table II to illustrate the relationship between baryon density and the ratio of proton fractions, $y_p^{\rm min}/y_p$, where $y_p$ is the  $\beta-$equilibrium proton fraction. The onset of the DU process occurs when this ratio, $y_p^{\rm min}/y_p$, is less  or equal to 1. By analyzing the onset of the DU process at $\rho_{dU}$, we can classify three sets of EOS based on similar densities: $\sim 2.5\rho_0$, $\sim 2.75\rho_0$ and $\sim 3.1\rho_0$ for $M_{DU}$ values of 1.6, 1.8, and 2.0$M_\odot$, respectively. As expected, the highest values of $M_{DU}$ correspond to the greatest values of $\rho_{DU}$, as shown in Table~\ref{tab3}. Since  the symmetry energy at saturation remains constant for both families, it is the parameter $L$ that governs the onset of DU processes. A larger value of $L$ results in a lower onset density, as a larger $L$ facilitates faster growth of symmetry energy beyond the saturation density.

\subsection{Symmetry energy and Mass-Radius \label{DD}}

The mass-radius relationship of a static NS can be determined by solving the Tolman-Oppenheimer-Volkoff (TOV) equation  \cite{TOV1,Oppenheimer:1939ne}, which incorporates the equation of state (EOS) of the NS, representing pressure as a function of energy density. We have included the Baym-Pethick-Sutherland outer crust \cite{Baym:1971pw}, and the inner crust was treated as in \cite{Malik:2022zol}: the outer crust is matched to the outer core EoS at $0.04$fm$^{-3}$ using a polytrope. The matching density was chosen in order to reduce the error introduced, which was shown to be below 100 m \cite{Malik:2024nva}.
Table ~\ref{tab3}  provides details on the maximum gravitational mass, its baryonic mass and radius, the mass $M_{\rm DU}$ and its central density $\rho_{\rm DU}$, as well as the radius and central density of stars with masses of 1.6, 1.8, and 2.0  $M_\odot$. While the properties of the maximum mass star remain largely unchanged with the modifications introduced in the isovector channel, there is a noticeable impact on the radius, central density, DU mass and radius of stars with masses less than or equal to 2.0$M_\odot$. Specifically, a larger slope $L$ results in smaller $M_{\rm DU}$ and density $\rho_{DU}$, leading to reduced NS radii and central densities. 

\begin{figure*}[tbh]
\includegraphics[width=0.9\linewidth]{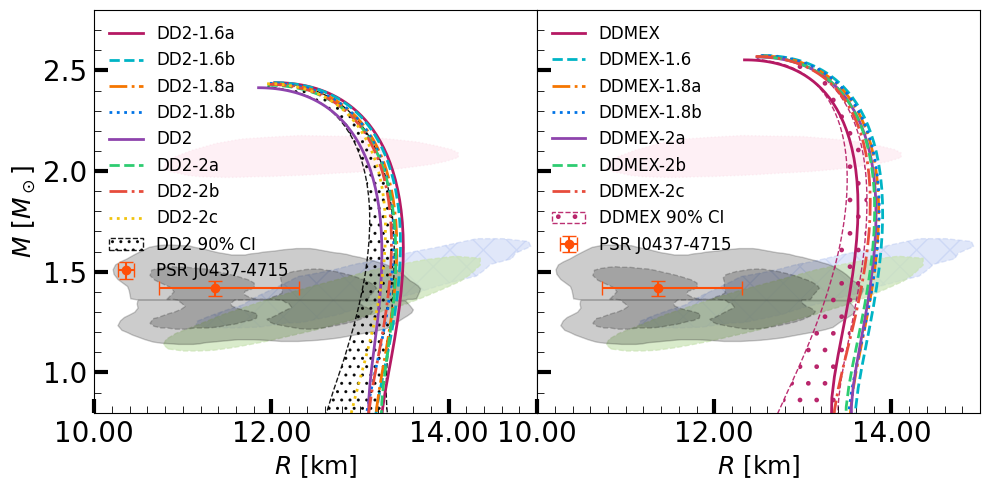}
\caption{Neutrons star masses versus the radius for the two DD families: left panel DD2 set, right panel DDMEX set. NS masses as a function of radius for the DD2 (blue) and DDMEX (orange) families, allowing the three couplings that define the isovector channel to vary, i.e. $\Gamma_\rho(\rho_0),\, a_\rho, \, y$. Both the median and the 90\% CI  probability regions are plotted. The LIGO-Virgo collaboration and NICER observations are also shown:  The grey lines represent the constraints derived from the binary components of GW170817, along with their corresponding 90\% and 50\% CI \cite{abbott2018}. Also shown are the $1 \sigma$ (68\%) CI for the 2D posterior distribution in the mass-radius domain for the millisecond pulsars PSR J0030 + 0451 (cyan and violet) \cite{Riley2019,Miller2019} and PSR J0740 + 6620 (purple and peach) \cite{Riley2021,Miller2021} from the NICER X-ray data.
}
\label{fig15}
\end{figure*}

In Fig.\ref{fig15}, we plot the mass-radius relation ($M-R$) for the modified DD2 and DDMEX models. Also included are some recent astrophysical observations (see the caption for details). As referred before, it is clearly seen that the maximum masses of NSs calculated using DD2 and DDMEX models are not very sensitive to the isovector channel, suffering a small increase not larger than 0.02$M_\odot$. Notably, the DDMEX models, also the generalized parametrizations,   demonstrate the capability to predict NSs exceeding 2.5$M_\odot$, consistent with the observed mass range of the secondary compact object in GW190814, reported as $2.50-2.67M_\odot$ (see \cite{Huang:2020cab}). %
It is clearly seen that the effect of $y$ and new values of $a_\rho$ that allow DU processes to occur inside NS is not negligible: the radius of  medium and high mass stars is 150-300 m larger due the much stiffer symmetry energy. 
In this figure we have also included the results of a Bayesian inference calculation that will be discussed below.

\begin{table*}[!ht]
\caption{The NS properties are given: the maximum mass $M_{\rm max}$ and the DU mass $M_{DU}$, respective baryon mass $M_{\rm B,max}$ and radius  $R_{\rm max}$, the radius of 1.4, 1.6, 1.8 and 2.0$M_\odot$ stars, the DU onset  baryon density $\rho_{DU}$ and the central baryon density $\rho_c$.}
\label{tab3}
\begin{ruledtabular}
\vspace{0.5cm}
\begin{tabular}{lccccccccc}
{\multirow{2}{*}{Model}} & $M_{\rm max}$& $M_{DU}$  & $M_{\rm B,max}$ & $R_{\rm max}$& $R_{1.6}$ &$R_{1.8}$& $R_{2.0}$& $\rho_{\rm DU}$ & $\rho_c$ \\ \cline{2-10} 
                          &\multicolumn{3}{c}{$M_\odot$} & \multicolumn{4}{c}{km}  & \multicolumn{2}{c}{fm$^{-3}$} \\ \hline
 \hline
DD2     & 2.42 & -- &2.92 & 11.87&13.22&13.21& 13.20 & -- & 0.852\\
DD2-1.6a & 2.44 & 1.6& 2.94 & 12.07&13.54&13.51&13.42 & $0.366$ & 0.827  \\
DD2-1.6b & 2.44 & 1.6& 2.94 & 12.06&13.50&13.48&13.40 & $ 0.37$&0.828  \\
DD2-1.8a  & 2.44 & 1.8& 2.94  & 12.02&13.44&13.43&13.33 & $0.417$ &0.833 \\
DD2-1.8b & 2.44 & 1.8& 2.94 & 12.02&13.40&13.39&13.31 & $0.415$  & 0.832 \\
DD2-2a& 2.43 & 2.0& 2.93  & 11.99&13.42&13.40&13.31   & $0.466$  & 0.838 \\
DD2-2b& 2.44& 2.0& 2.94  & 11.98&13.36&13.35&13.27   & $0.472$& 0.836  \\
DD2-2c& 2.44 & 2.0& 2.94  & 11.97&13.28&13.28&13.22 & $0.464$  &0.834 \\
\hline
DDMEX     & 2.56 & -- & 3.11 & 12.35 &13.46&13.51&13.53&--&0.776  \\
DDMEX-1.6  & 2.58 & 1.6 & 3.13&  12.56&13.87&13.90&13.87   & $0.343$& 0.756\\
DDMEX-1.8a& 2.58 & 1.8 &3.13& 12.53&13.80&13.84&13.83& $0.366$&  0.758\\
DDMEX-1.8b &2.58 & 1.8 &3.13 & 12.54&13.81&13.86&13.83 & $0.357$& 0.758 \\
DDMEX-2a  & 2.57  & 2.0 & 3.12 & 12.49 &13.76&13.80&13.78 & $0.40$& 0.763\\
DDMEX-2b & 2.57  & 2.0 & 3.12& 12.49 &13.74&13.78&13.77 & $0.397$& 0.763 \\
DDMEX-2c  & 2.57 & 2.0 & 3.13& 12.47 &13.67&13.73&13.72 & $0.403$& 0.763 \\
\end{tabular}
\end{ruledtabular}
\end{table*}
In addition to the constraints provided by observations of massive NSs such as PSR J1614-2230, PSR J034+0432, and PSR J0740+6620, recent measurements from NICER have simultaneously determined the mass and radius of a NS in the intermediate mass region, PSR J0030+0451. According to~\cite{Miller2019} its mass is 1.44$_{-0.14}^{+0.15}$M$_\odot$ with a radius of $13.02_{-1.06}^{+1.24}$km, while \cite{Riley2019} reports a mass of $1.34_{-0.16}^{+0.15}~M_\odot$ with a radius of $12.71_{-1.19}^{+1.14}$ km. Across various models, as verified in  previous studies \cite{Carriere:2002bx,Cavagnoli11,Providencia:2012rx}, it is noted that a larger slope corresponds to larger radii for NSs with masses less than or equal to 2.0M$_\odot$. Specifically, for 1.4M$_\odot$ stars, the radius increases from 13 km to 13.5 km.

\subsection{The isovector channel: Bayesian inference \label{sec2d}}

In the previous section we discussed the main properties of the DD2 and DDMEX families, keeping fixed the isoscalar channel describing the symmetric nuclear matter EOS and the symmetry energy at saturation: the variation of the parameters $y$ and $a_\rho$ has allowed to vary the slope and the curvature of the symmetry energy at saturation. As discussed, a direct consequence was the effect these two parameters have on the onset of the nucleonic DU processes, allowing a decrease in the onset density and consequently a decrease in the mass of the NS with a central density coinciding with the DU onset density.

\begin{figure*}[ht]
\begin{center}
\begin{tabular}{c}
\includegraphics[angle=0, width=1.0\linewidth]{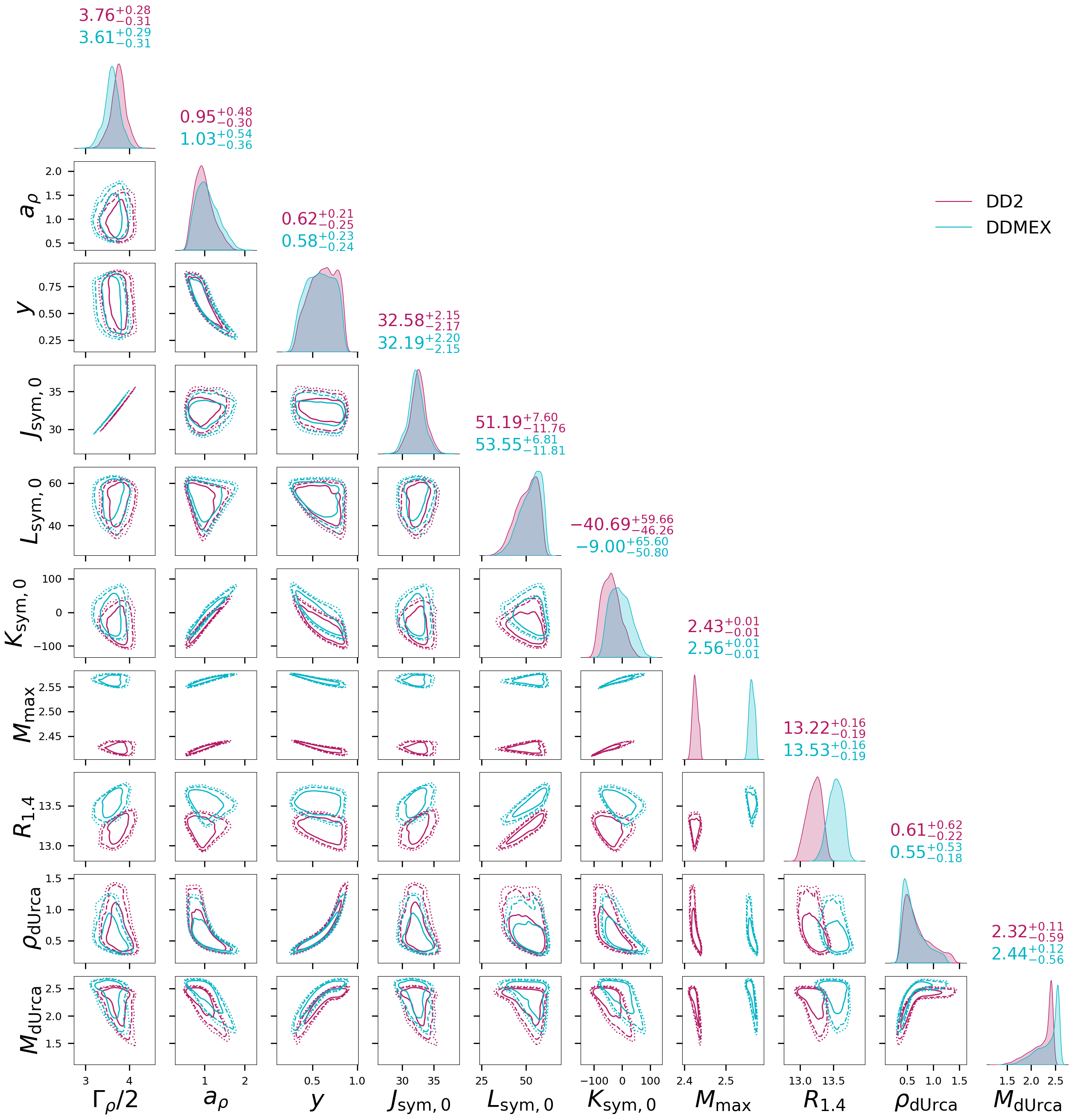}\\
\end{tabular}
\end{center}
\caption{Corner plot obtained by varying the parameters responsible for the isovector channel, $\Gamma_{\rho_0}$, $a_{\rho}$ and $y$, using the parameterisation of DD2 (pink-red) and DDMEX (light-blue) for the isoscalar channel. Isovector nuclear matter properties ($J_{\rm sym, 0}\equiv E_{\rm sym, 0}$, $L_{\rm sym, 0}\equiv L_{0}$, $K_{\rm sym, 0}$) and NS properties (the maximum mass $M_{\rm max}$, the radius for 1. 4 $M_\odot$ NS $R_{1.4}$, the onset density of the DU processes $\rho_{\rm DU}$ and the corresponding NS mass with this density at the center $M_{\rm DU}$) are shown. The different line styles in the two-dimensional distribution (solid, dashed and dotted) correspond to 68\%, 95\% and 99\% CI.}
\label{corner1}
\end{figure*}

In the following we present a Bayesian analysis to constrain the complete isovector channel, i.e. the couplings $\Gamma_\rho(\rho_0),\, a_\rho$ and $y$, considering as constraints the onset of nucleonic  DU and the low density chEFT neutron matter pressure.
The corner plot shown in Fig. \ref{corner1} summarises the main results: The bidimensional distributions relate the isovector channel parameters to the isovector nuclear matter properties $J_{\rm sym, 0}\equiv E_{\rm sym, 0}$, $L_{0}$, $K_{\rm sym, 0}$, and to the NS properties maximum mass $M_{\rm max}$, radius for 1.4~$M_\odot$ NS $R_{1.4}$, onset density of the DU processes $\rho_{\rm DU}$ and corresponding NS mass with this density at the center $M_{\rm DU}$. The two models previously discussed, DD2 and DDMEX, were considered as a starting point, in particular keeping all the parameters defining the isoscalar channel fixed. The results for the two models are shown in different colours, pink-red for DD2 and light-blue for DDMEX.

\begin{figure*}[tbh]
    \centering
    \includegraphics[width=0.85\linewidth]{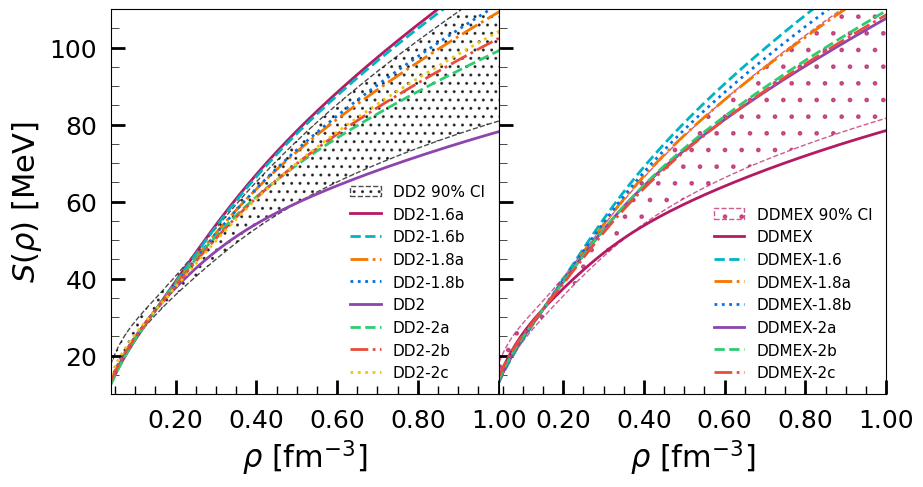}
    \caption{The symmetry energy, denoted as $S(\rho)\equiv E_{\rm sym}(\rho_{B})$, is plotted against the number density $\rho$. Both $S(\rho)$ and $\rho$ are measured in MeV and fm$^{-3}$ respectively, across various isovector variations of the DD2 and DDMEX models. The median curve is depicted together with the 90\% confidence interval.
    }
    \label{fig:symeng}
\end{figure*}

The main conclusions that can be drawn are:
1) the parameter $y$ correlates not only with the onset density of DU, and corresponding mass, but also with the maximum star mass. The smaller $y$ the smaller the onset density of DU and the larger the maximum star mass. Considering the constraints imposed it was not possible to get values for the mass $M_{DU}$ below $\sim2M_\odot$ for the DDMEX family, but it is possible to go down to $\sim 1.6\, M_\odot$ for the DD2 family at 99\% CI. The parameter $y$ also affects the properties $L_{0}$, $K_{\rm sym, 0}$, in particular, a smaller value of $y$ pushes these two quantities to larger values; 2) $a_\rho$ correlates strongly with the curvature $K_{sym,0}$, giving larger values, that can even be  positive, for larger  values of $a_\rho$. $K_{sym,0}$ is in average larger for the DDMEX family, and at 68\% CI it can take values of the order $\sim +30$~MeV. The largest values of $M_{max}$ occur for the largest values of $a_\rho$; 3) there is a clear correlation between $L_{0}$ and $R_{1.4}$ as shown in other studies when the isoscalar channel is kept fixed \cite{Carriere:2002bx, Cavagnoli11,Providencia:2012rx,Salinas:2023nci}. Notice, however, that if the isoscalar channel is also allowed to vary this correlation seems to disappear \cite{Alam:2016cli,Char:2023fue}. For both models chEFT constraints impose $40\lesssim L_{0}\lesssim 60$ MeV; 4) The coupling $\Gamma_{\rho_0}$, is totally correlated with $E_{sym,0}$ by definition, and also correlated with $K_{sym,0}$ the smaller values of this properties being associated with the larger values of the coupling $\Gamma_{\rho_0}$;
5) the isovector channel does not affect much the NS maximum mass, a maximum variation of 0.1$M_\odot$ is obtained. It is also observed that the DDMEX family predicts  masses $\sim 0.2\, M_\odot$ larger than the DD2 family; 6) finally let us also point down the behavior of the families concerning the radius of the 1.4$M\odot$, $R_{1.4}$: in average DD2 predicts radii $\sim 400$m smaller than DDMEX family; 
 7) notice that no clear correlation between $L_{0}$ and $K_{\rm sym, 0}$ is obtained, as discussed in the previous sections.

The comparison of the above results for the NS properties with the previous discussion  allows some interesting conclusions:
 the prediction of the M$_{DU}$ obtained are compatible with the values calculated in Table ~\ref{tab3}, the lowest $M_{\rm DU}$  obtained was 1.6~M$_\odot$ for the DD2 family with a slope $\rm L_{0}$ greater than $60$~MeV with $68\%$CI and $\rm K_{sym,0}$ slightly lower than 0, see DD2-1.6a, DD2-1.6b models in Tables~\ref{Tab11} and~\ref{tab3}. The models DD2-1.8a, DD2-1.8b with a $M_{\rm DU}=1.8M_\odot$ have   a slope below $60$ MeV. For DD2-2b, DD2-2c with $M_{\rm DU}=2.0M_\odot$ with a slope around $55$MeV (for $y>0.4$, $a_\rho$ around 1.0) and a negative value of $K_{\rm sym, 0}\lesssim 0$ and the density $\rho_{\rm DU}$ is between $0.37$~fm$^{-3}$ and $0.48$~fm$^{-3}$. Note, however, that DD2-2a has a slope above 60 MeV because $y$ takes a value above 0.5 contrary to all the other DD2 parametrizations.  In summary, the smallest $M_{\rm DU}$  requires the largest values of $L_{0}$ and values of $K_{\rm sym, 0}$ close to zero.  However,  large values of  $M_{\rm DU}$  may also have large values of $L_{0}$, but in this case the $K_{\rm sym, 0}$ parameter is more negative: the larger $L_{0}$ the more negative $K_{\rm sym, 0}$.

For reference we show the 90\% CI distribution of the symmetry energy in Fig. \ref{fig:symeng}. Also included, for comparison are the respective curves for DD2 and DDMEX. As expected, the DU constrain imposes a much harder symmetry energy at high densities.

In Fig. \ref{fig15}, the mass-radius 90\% CI distribution resulting from the inference calculation is plotted for both DD2 and DDMEX families.   These distributions have a large overlap for $M\lesssim 1.4 \, M_\odot$, because this is the region that is most affected by the  chEFT constraints. Both models are compatible with the present observations due to the large uncertainties still associated with the measurements. Note that some of the parametrizations proposed in Table \ref{tab3} lie outside the 90\% credible level distributions because only extreme parametrizations could satisfy the conditions necessary to allow DU processes inside the NS.

\subsection{Comparing models through the Bayes factor}

In the last subsection, we have determined the neutron star EOS space spanned by the three couplings responsible for the isovector channel through a Bayesian inference calculation, subject to a minimal set of neutron star and nuclear matter constraints, and imposing the condition that the direct Urca processes involving nucleons occur inside neutron stars. This is a condition that is not satisfied by the DD2 and DDMEX parametrizations. 

We recall that in the first subsection of this section, subsection \ref{DD}, we derived a set of models that predict DU at 1.6, 1.8 and 2.0 $M_\odot$, keeping the symmetry energy at saturation equal to that of the original models DD2 and DDMEX, thus with an additional condition regarding the inference computation. We can now ask, what is the likelihood of these models determined outside of Bayesian inference?

\begin{table}[htbp]
\centering
\caption{The DU mass,  the log of the evidence and log of the  Bayes factor with respect to DD2B and DDMEXB, for the EOS with the largest likelihood of the two sets, respectively,  DD2B and DDMEXB, and for the  EOS with largest likelihood for a given DU mass in each set, respectively DD2B-M$_{DU}$ and DDMEXB-M$_{DU}$.
The $\mathcal{B}_{M}^{i}$ = ln $\mathcal{Z}_{\rm tot}^{\rm EOS}$/ln $\mathcal{Z}_{\rm tot}^{xB}$ with $x=$DD2, DDMEX.}
\begin{tabular}{lccccc}
\hline 
\hline 
EOS & $M_{\rm dUrca}$ & ln $\mathcal{Z}_{\rm tot}$ & ln $\mathcal{Z}_{\rm NMP}$ & ln $\mathcal{Z}_{\rm chEFT}$ & ln $\mathcal{B}_{M}^{i}$ \\
\midrule
DD2B      & 2.351 & -1.99 & -5.55 & 3.56 & 0.00  \\
DD2B-1.6  & 1.606 & -2.28 & -5.69 & 3.42 & -0.29 \\
DD2B-1.8  & 1.808 & -2.12 & -5.63 & 3.51 & -0.13 \\
DD2B-2.0  & 2.013 & -2.03 & -5.57 & 3.54 & -0.04 \\
\\ 
DDMEXB    & 2.239 & -2.29 & -5.85 & 3.56 & 0.00  \\
DDMEXB-1.6& 1.600 & -2.90 & -6.40 & 3.50 & -0.61 \\
DDMEXB-1.8& 1.796 & -2.40 & -5.94 & 3.54 & -0.11 \\
DDMEXB-2.0& 2.014 & -2.31 & -5.86 & 3.56 & -0.01 \\
\bottomrule
\end{tabular}
\label{lnBin}
\end{table}

In order to answer this question we have determined for each inference model, DD2 inference model and DDMEX inference model, the EOS with highest likelihood, i.e. the largest Bayesian evidence,  and took these two EOS as reference. They will be designated by DD2B and DDMEXB, respectively. The properties of both EOS are given in the Appendix, Table \ref{tab:a1} and Fig. \ref{fig:a1}. Note that they predict DU inside NS by construction, but only for masses above 2.3~$M_\odot$.  In addition to these two EOS, we have also looked in each inference model for the three EOS with highest likelihood  predicting DU processes inside NS with 1.6, 1.8 and 2.0 $M_\odot$, see Tables \ref{lnBin} and \ref{tab:a1}, respectively, for the Bayes factors with respect to the models DD2B and DDMEXB and the nuclear matter and NS properties. In Table \ref{lnBin}, the  third column defines the likelihood of each model (the logarithm of the Bayesian evidence), and the  fourth  columns gives the  logarithm of the Bayes factor with respect to DD2B or DDMEXB. The Bayes factors indicate that: i) the larger the DU mass the larger the likelihood. This is not surprising because the EOS DD2B and DDMEXB predict quite large DU masses, respectively 2.35$M_\odot$ and 2.33$M_\odot$; ii)  the absolute  magnitude of the  Bayes factors  is below 1 indicating that the likelihood of the three EOS within each inference model do not differ much from the likelihood of  DD2B and DDMEXB  EOS.  These EOS will allow us to assess the likelihood of the EOS given in Table \ref{tab3}.

It is also interesting to compare the isovector  properties of these EOS with the ones of the EOS in Table \ref{tab3}. All DD2Bx and DDMEXBx  EOS have a larger symmetry energy at saturation than DD2 and DDMEX, and than all models introduced in subsection \ref{DD} which  have a symmetry energy equal to the one of DD2 or DDMEX  by construction. The smaller the DU mass the larger the symmetry energy at saturation. The compatibility with chEFT constraints is attained with smaller values of the slope and larger values of the curvature of the symmetry energy at saturation. In particular, the curvature $K_{sym}$ takes always positive values for DDMEXBx if DU mass is equal or below 2$M_\odot$  (and for DD2Bx if DU is smaller than 2 $M_\odot$) and may be as high as 67 MeV. In \cite{Li:2021thg} a range $K_{sym} = -107 \pm 88$ MeV is proposed at a 68\% confidence level. Note, however, that parametrizations with a quite high value of $K_{sym}$ have been proposed in \cite{Reed2023}.

\begin{table}[htbp]
\setlength{\tabcolsep}{5.0pt}
      \renewcommand{\arraystretch}{1.1}
\centering
\caption{Logarithm of the total evidence ln $\mathcal{Z}_{\rm tot}$, and of  the NMP and chEFT contributions to the evidence (
ln $\mathcal{Z}_{\rm NMP}$ and ln $\mathcal{Z}_{\rm chEFT}$) of the EOSs  given in Table \ref{tab3}. The last column defines the logarithm of the Bayes factor with respect to the DD2B and DDMEXB EOS, 
ln $\mathcal{B}^i_M$ = ln $\mathcal{Z}_{\rm tot}^i$/ln $\mathcal{Z}_{\rm tot}^{xB}$ with $x=$DD2, DDMEX. Based on Jeffreys' scale of Bayes factors, the categories Decisive, Strong, and Substantial correspond to units above 5.0, 2.5, and 1.0, respectively, on the natural logarithm scale \cite{jeffreys1998theory, Morey_2016}. Here, the Bayes factor is computed with respect to the optimal model in the DD2 posterior for the DD2 family and the DDMEX posterior for the DDMEX family, respectively DD2B and DDMEXB.  The models with ln $\mathcal{B}^i_M$ greater than -1 are on par with the optimal model.}
\begin{tabular}{ l c c c c}
\hline \hline 
EOS       & ln $\mathcal{Z}_{\rm tot}$      & ln $\mathcal{Z}_{\rm NMP}$   & ln $\mathcal{Z}_{\rm chEFT}$   &  ln $\mathcal{B}^i_M$  \\
\midrule
DD2          & -2.60  & -5.67  & 3.06  & -0.61 \\
DD2-1.6a     & -10.46 & -5.67  & -4.79 & -8.46 \\
DD2-1.6b     & -6.18  & -5.67  & -0.51 & -4.19 \\
DD2-1.8a     & -2.65  & -5.67  & 3.01  & -0.66 \\
DD2-1.8b     & -2.34  & -5.67  & 3.33  & -0.34 \\
DD2-2a       & -3.48  & -5.67  & 2.19  & -1.48 \\
DD2-2b       & -2.17  & -5.67  & 3.50  & -0.17 \\
DD2-2c       & -2.55  & -5.67  & 3.12  & -0.56 \\
\\ 
DDMEX        & -2.43  & -5.86  & 3.44  & -0.13 \\
DDMEX-1.6    & -17.27 & -5.86  & -11.41 & -14.98 \\
DDMEX-1.8a   & -9.73  & -5.86  & -3.87  & -7.44 \\
DDMEX-1.8b   & -11.24 & -5.86  & -5.37  & -8.94 \\
DDMEX-2a     & -5.88  & -5.86  & -0.02  & -3.59 \\
DDMEX-2b     & -5.50  & -5.86  & 0.36   & -3.21 \\
DDMEX-2c     & -4.82  & -5.86  & 1.04   & -2.53 \\ 
\hline 
\end{tabular}
\label{tab:lnB}
\end{table}

 We next analyse the likelihood of the EOS defined in subsection \ref{DD} with the properties given in Table \ref{tab3}.  In Table \ref{tab:lnB}, we present for all these EOS the logarithm of the total evidence (first column), the terms corresponding to the NMP and the chEFT constraints (second and third columns) and  in the last column the Bayes factor with respect to the EOS DD2B and DDMEXB. Some comments are in order:  i) it is the chEFT condition that gives origin to the different evidence within each model (DD2 or DDMEX); ii) as expected, the EOSs with a smaller Bayes factor are those with smaller DU masses, since these are the ones that are less compatible with the chEFT;  iii) regarding the DD2 family, EOS DD2-1.6a and b have, respectively,  decisive and strong evidence against them, DD2-2a  has a substantial evidence against it,  and all the other EOS show an evidence similar to that of DD2B; iii) For the DDMEX family, all EOSs are quite disfavored. All with a DU mass equal to or less than 1.8$M_\odot$ are decisively disfavored, and those with M$_{DU}=2\, M_\odot$ are strongly disfavored.   These results indicate that the DD2 EOS of table \ref{tab3} have a similar likelihood to the reference EOS, except for the EOS with a DU mass equal to 1.6$M_\odot$, but the DDMEX EOS are quite disfavored due to the incompatibility with the chEFT constraints. The models in Table \ref{tab3} have the particularity of keeping the original parameter $\Gamma_{\rho}(\rho_0)$ of both models DD2 and DDMEX. As a consequence, the symmetry energy at saturation remains unchanged, respectively 30.8 and 32.3 MeV, well within the range predicted within a chEFT calculation \cite{Drischler2020}, 31.7$\pm1.1$ MeV. Models in Table \ref{lnBin} all have a larger coupling  $\Gamma_{\rho}(\rho_0)$, and, therefore a larger symmetry energy at saturation  that can take values above 34 MeV, see Table \ref{tab:a1}.  The other difference already pointed out is that DD2Bx and DDMEXBx take larger values, generally positive, for the symmetry energy curvature $K_{sym}$.

\section{Conclusion \label{concl}}
In order to study the properties of NSs, in this paper, we have considered two families of density-dependent relativistic mean-field models with a stiff EOS, namely the DD2 and DDMEX families. These models are able to generate NSs with masses as large as  2.44~M$_\odot$ (DD2) and 2.57~M$_\odot$ (DDMEX). In particular, DDMEX was  constrained by the low mass compact object in the event GW190814, with a mass of 2.50-2.67 ~M$_\odot$ \cite{Huang:2020cab}.
A common behavior of many DD models is that they do not predict the nucleonic DU processes inside nucleonic NSs (\cite{Fortin2016,Fortin:2021umb}), due to the exponential decrease of the $\rho$-meson coupling with density. To explain the cooling curves of the thermal evolution of isolated non-magnetized and non-rotating NSs or accreting NSs with these models, \cite{Providencia:2018ywl,Fortin:2021umb} found it necessary to include hyperons inside the star. This may have the consequence of making the EOS too soft, and, therefore, not allowing for the existence of 2$M_\odot$ mass stars, if the hadronic EOS is soft. 

To overcome this limitation, we adopt a different approach and introduce  a generalization of the $\rho$-meson coupling constrained by observations of NS cooling as shown in \cite{Malik2022}, allowing for the nucleonic DU processes to occur inside NS. The objective of the present work was to investigate the consequence of this new parameter on the nuclear matter properties and NS properties.
By adopting as a  high-density constraint, the prediction of nucleonic DU processes inside NS, we have generated a set of DD models that obey the $\chi$EFT EOS neutron matter constraints and satisfy  nuclear matter saturation properties, namely saturation density, binding energy per particle, incomprehensibility and symmetry energy, according to the presently accepted values. For the isospin channel at low densities,  we have  considered an interval compatible with $\chi$EFT calculations. In addition, the symmetry energy of the models was also constrained in order to produce a maximum star mass  of at least 2~$M_\odot$. Notice, however, that the isospin channel does not have a large influence on the maximum mass and we did not obtain deviation larger than 0.1$M_\odot$ from the original maximum masses predicted by DD2 and DDMEX.

The models we have obtained allowed us to predict an onset of DU cooling processes inside stars with a mass $\gtrsim$1.6~$M_\odot$ ($\gtrsim$1.8~$M_\odot$) for the DD2 (DDMEX) family with maximum masses and radii slightly larger than the maximum M-R of the original DD models ($2.42M_\odot$ for DD2 and $2.56M_\odot$ for DDMEX) and could still be compatible with the low-mass object of the binary merger at the origin of GW190814, 2.50-2.67~$M_\odot$ \cite{abbott2020b}. We note that considering the isoscalar channels of the models DD2 and DDMEX, the chEFT constraints excludes the appearance of the DU processes inside NS of masses below  1.6~$M_\odot$.  Within these models, if stars with masses below 1.6$M_\odot$ show a fast cooling it will be necessary to include hyperons, which could be envisaged in future work.

In order to confirm our results of subsection \ref{DD}, we have performed two inference calculations based on the DD2 and DDMEX models, and allowed the three parameters that characterize the isovector channel to vary, imposing that nucleonic DU processes are allowed inside NS. This allowed us to study the effect on the NMP and NS properties.  Using these two inference models as reference we have determined the EOS with the highest likelihood in each inference model, respectively EOSs  DD2B and DDMEXB. The Bayes factors of the EOSs proposed in Table \ref{tab3} were calculated with respect to these two  high likelihood EOS and it was  found that for the DD2 family all, except for the EOSs predicting DU inside 1.6 $M_\odot$ stars,  a likelihood similar to the reference model DD2B.  For the DDMEX family all models proposed are strongly or decisively disfavored with respect to DDMEXB, the main reason being the incompatibility with the chEFT constraints considered in our study. These models, however, where built keeping the symmetry energy at saturation equal to the one of the original models, and have smaller values of the symmetry energy and its curvature at saturation, while still presenting values for the slope of the symmetry energy within in acceptable range considering the values proposed in \cite{Lattimer:2023rpe}, $L=53\pm13$ MeV, at one $\sigma$.

We have shown that to decrease the DU mass, $M_{\rm DU}$, it is necessary to increase the symmetry energy slope at saturation, $L_0$. A direct consequence on the NS properties of increasing $L_0$ keeping the properties of the isoscalar channel fixed is the increase of the NS radius, the smaller the mass the larger the effect. At the maximum mass the effect was small but there was a clear reduction of the central density $\rho_c$. 

It was shown that in order to allow for DU nucleonic processes inside NS described by these two families of EOS, $K_{\rm sym, 0}$ may take values close to zero or positive. 

In the literature it is often pointed out a correlation between $L_0$ and $K_{\rm sym, 0}$  \cite{Providencia14,Ducoin2010,Ducoin2011}. However, for the modified DD EOS verifying  the high density constraints imposed by the DU process, the correlation $L_0$-$K_{\rm sym, 0}$ is  washed out for both families of EOS. It  is interesting to understand which are the effects on the low density region, below the saturation density, of considering the $L_0$-$K_{\rm sym, 0}$ coefficients determined from the high density constraints. These affect the densities that define the  crust  of the NS.  More specifically, we are interested in studying the impact of the high-density constrained DD models on the critical and crust-core transitions properties.

\section*{Acknowledgments}
 This research received partial funding from national sources through FCT (Fundação para a Ciência e a Tecnologia, I.P, Portugal) for projects UIDB/04564/2020 and UIDP/04564/2020, identified by DOI 10.54499/UIDB/04564/2020 and 10.54499/UIDP/04564/2020, respectively, as well as for project 2022.06460.PTDC with DOI 10.54499/2022.06460.PTDC. The authors acknowledge the Laboratory for Advanced Computing at the University of Coimbra for providing {HPC} resources that have contributed to the research results reported within this paper, URL: \hyperlink{https://www.uc.pt/lca}{https://www.uc.pt/lca}.

\section*{Appendix} 
\begin{figure*}[h]
\begin{center}
\hspace*{-1.0cm}
\begin{tabular}{cc}
\includegraphics[width=0.54\linewidth]{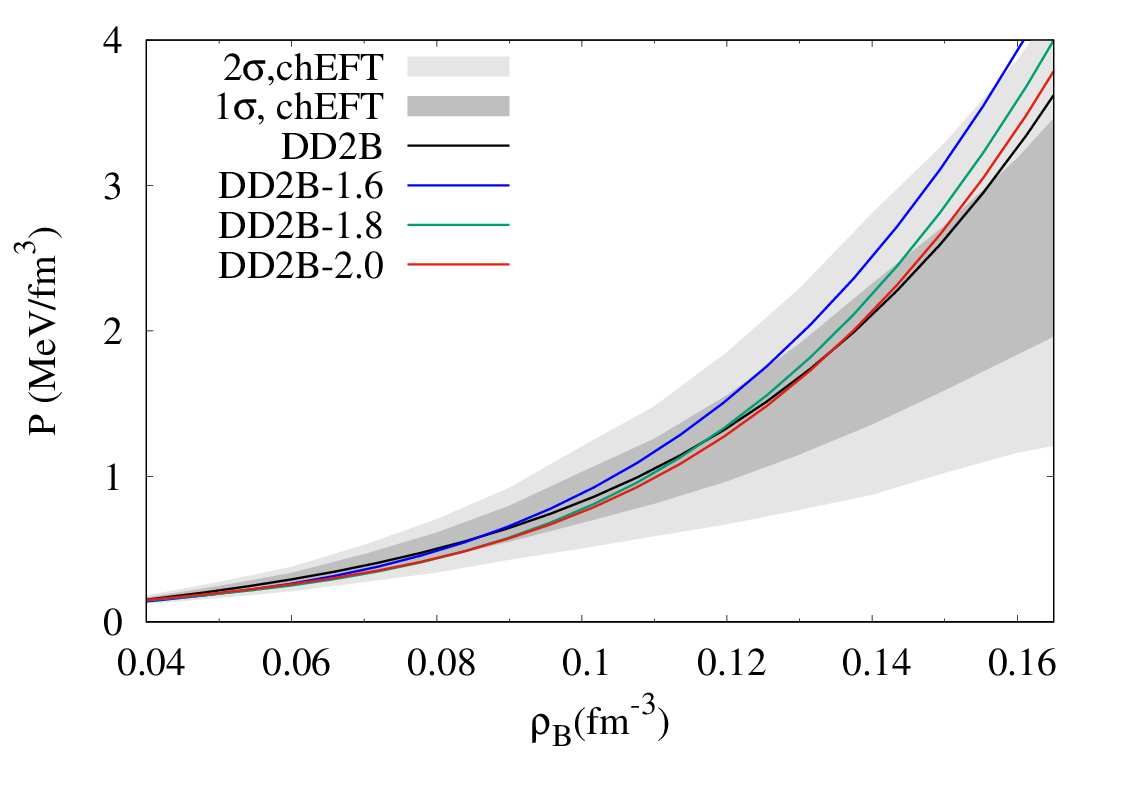}&
\includegraphics[width=0.54\linewidth]{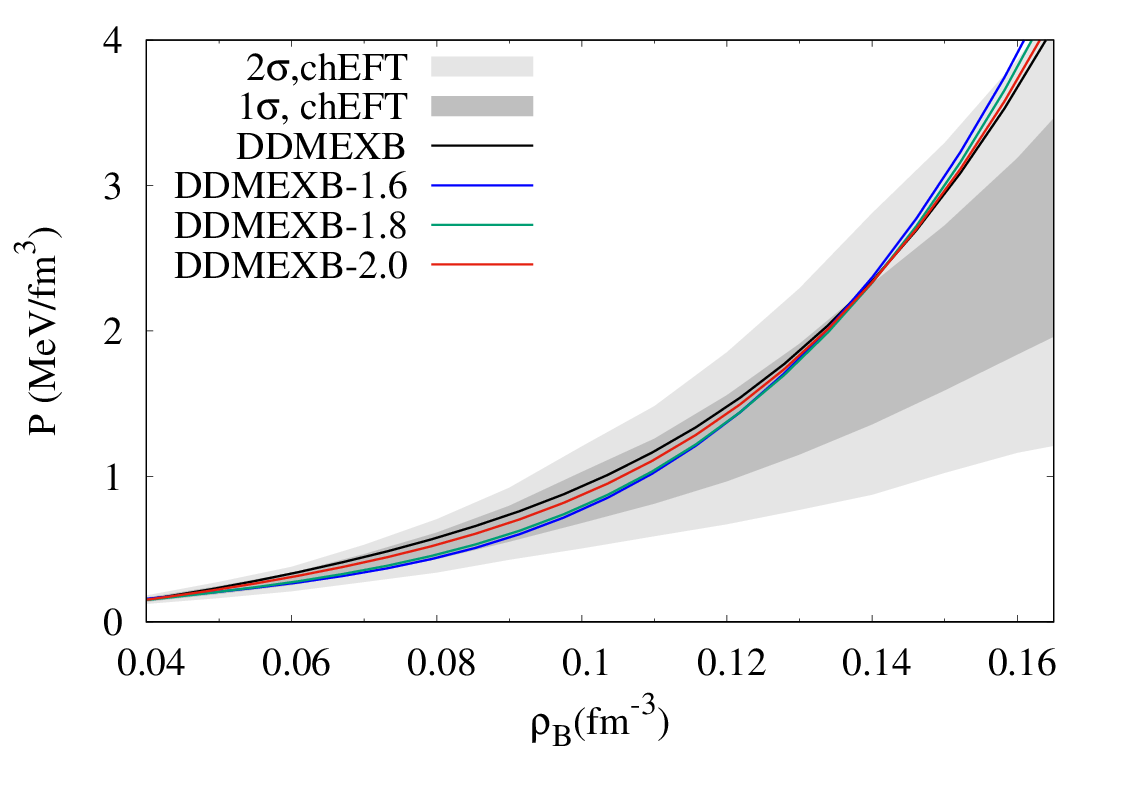}\\
\includegraphics[width=0.50\linewidth]{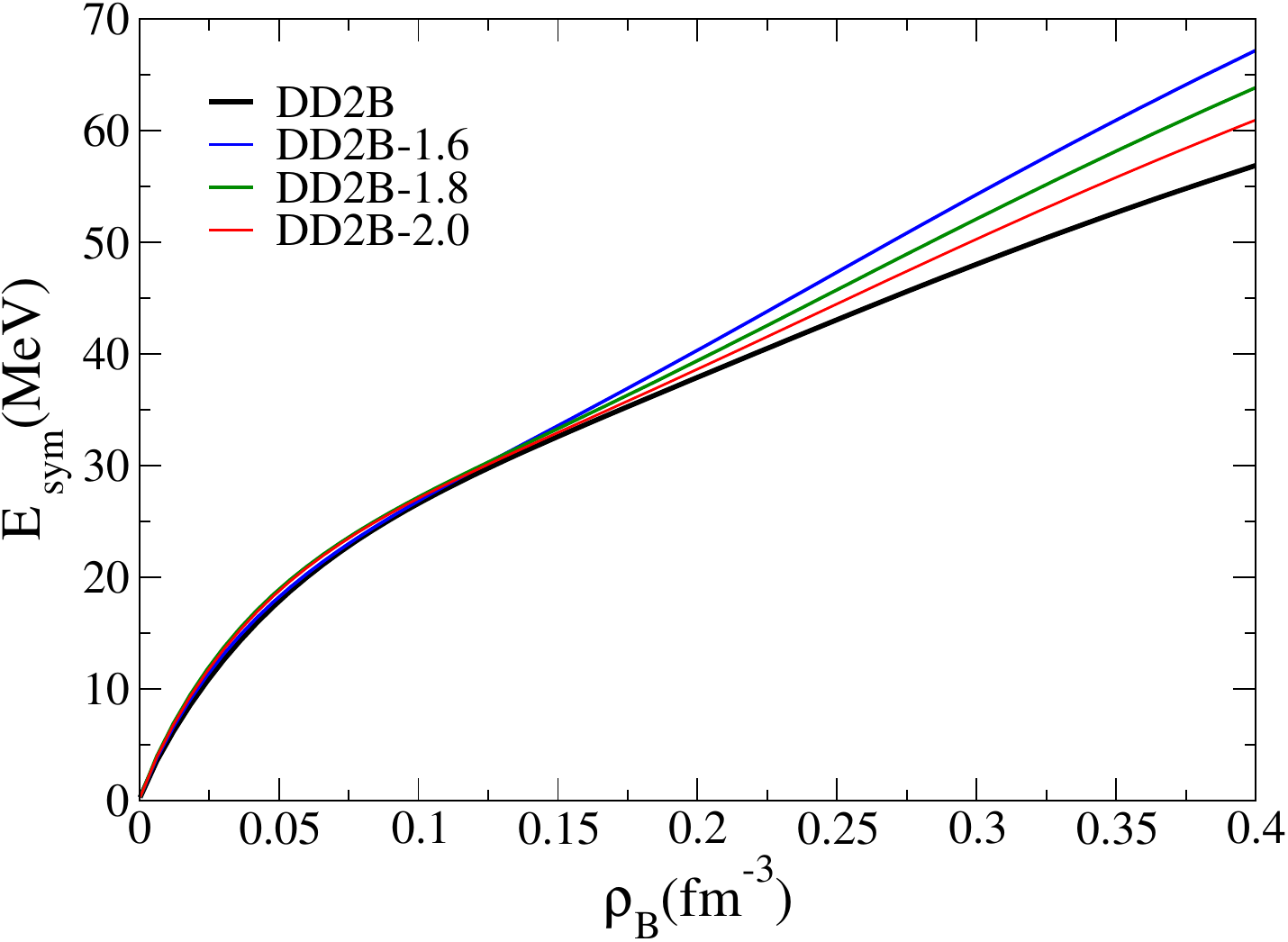}&
\includegraphics[width=0.50\linewidth]{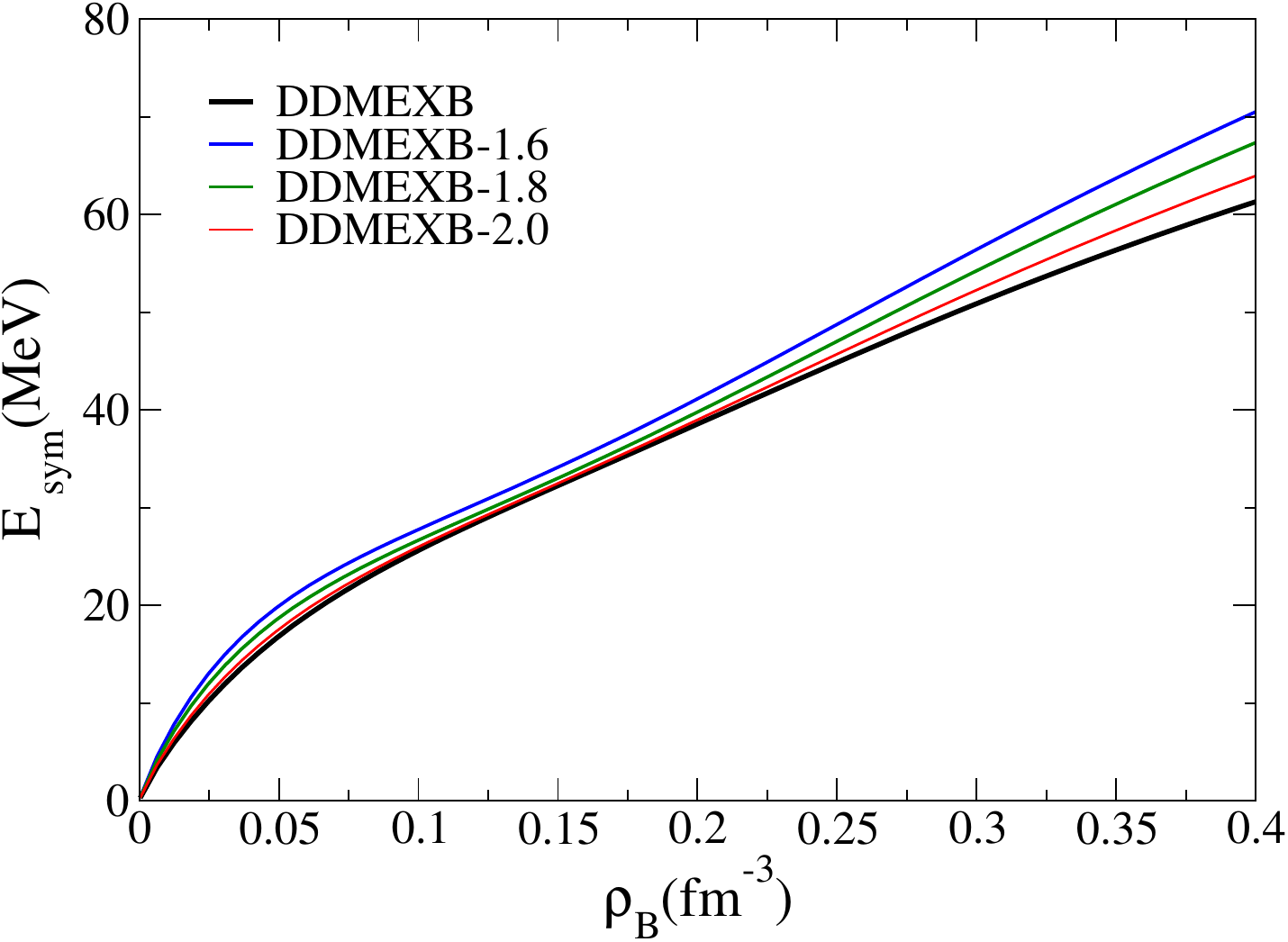}\\
\includegraphics[width=0.50\linewidth]{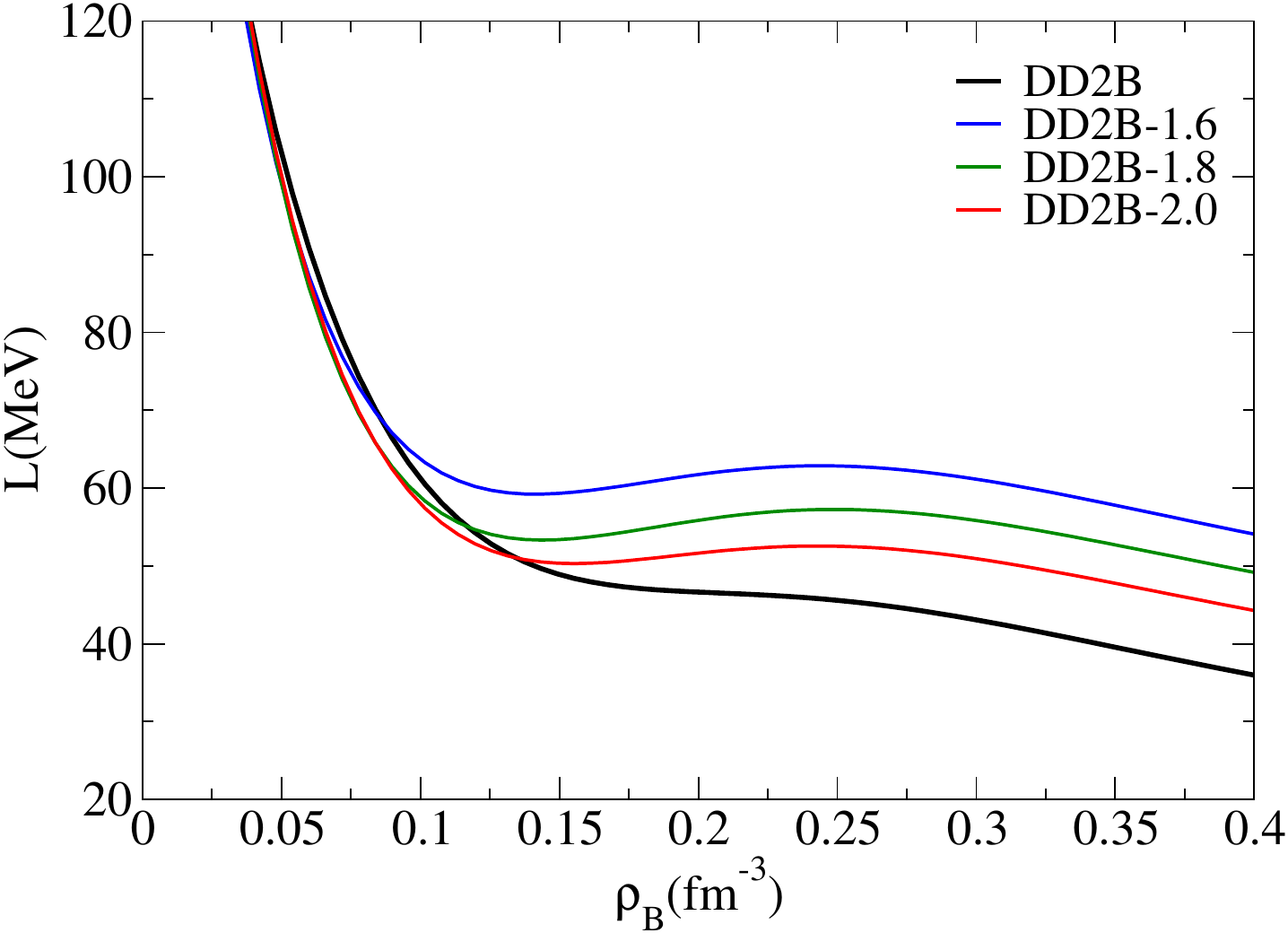}&
\includegraphics[width=0.50\linewidth]{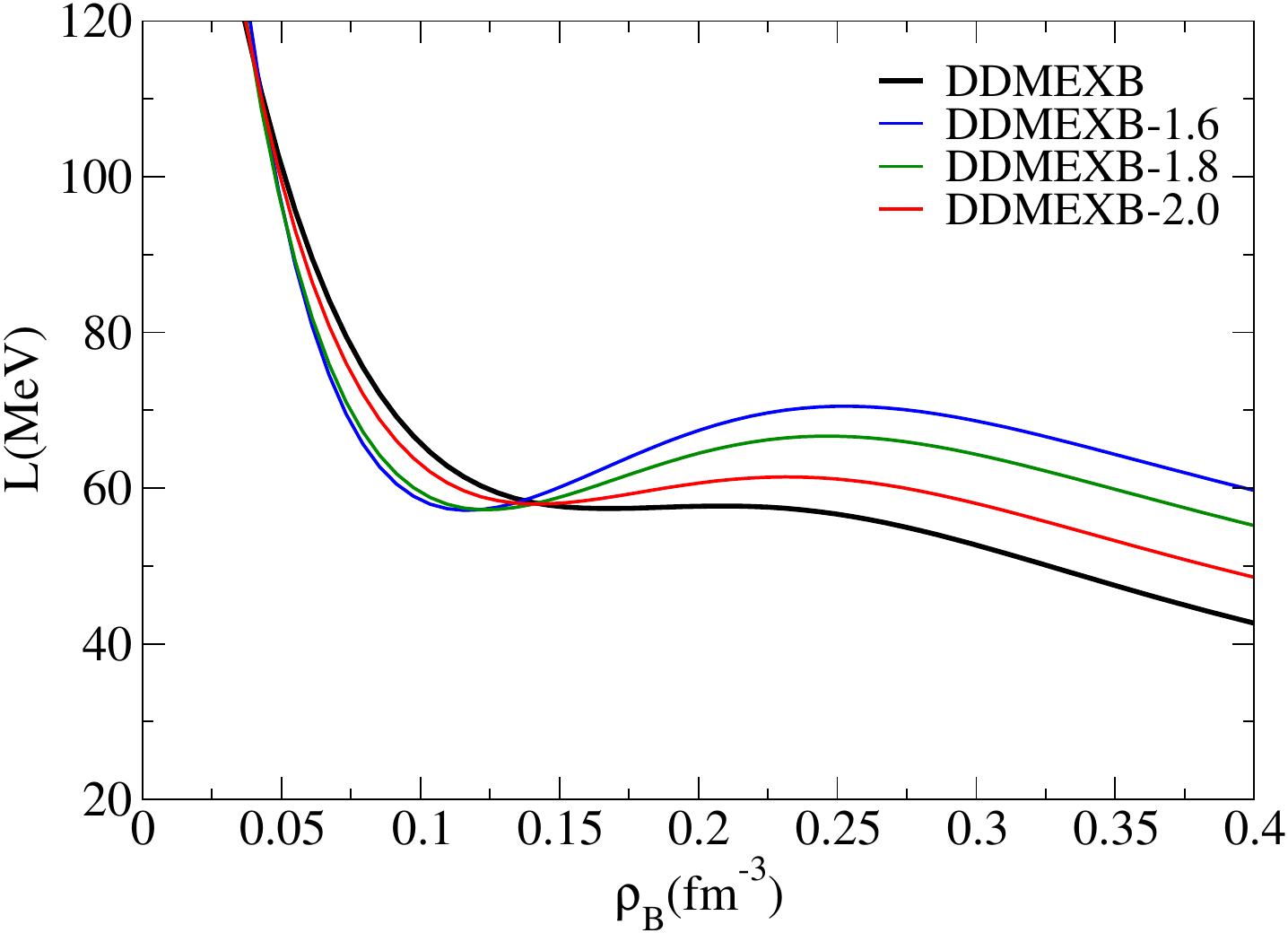}
\end{tabular} 
\caption{(Color online) Upper panels show the pure neutron matter pressure as a function of the baryonic density for DD2B (left) and DDMEXB (right) for the optimal model of families under consideration featuring urca masses of 1.6, 1.8, and 2.0 M$_\odot$ for both the DD2 and DDMEX families and including the 1$\sigma$ and 2$\sigma$ bands from chiral effective field theoretical calculations \cite{Hebeler2013}. Middle panels reveal the symmetry energy and in the bottom its slope as function of the baryon density for  the optimal model of DD2B (left) and DDMEXB (right) families.}
\label{fig:a1}
\end{center}
\end{figure*}

We present in Table \ref{tab:a1}  the isovector parameters, $\Gamma_\rho$, $a_\rho$ and $y$,  the NS and NMP properties of the  EOSs within each inference set, DD2 and DDMEX, with the highest likelihood in the posterior sets, DD2B and DDMEXB, and the three EOS with the highest likelihood of predicting nucleonic DU processes inside stars with masses of 1.6, 1.8, and 2.0 $M_\odot$, respectively. For reference, the  same nuclear matter properties shown in Fig. \ref{fig13} are also shown for the DD2Bx and DDMEXBx EOS in Fig. \ref{fig:a1}. 

\begin{table*}[h!]
\caption{We present a table summarizing the isovector parameters for the optimal model, which exhibits the highest total log-likelihood (ln $\mathcal{Z}_{\rm tot}$), along with a selection of models featuring Urca masses of 1.6, 1.8, and 2.0 M$_\odot$ for both the DD2 and DDMEX families. These selected models are chosen for their close resemblance to the best model within their respective families, characterized by a minimal Bayes factor ln $\mathcal{B}^i_M$ relative to the optimal model. Additionally, we provide the various slope parameters for the symmetry energy, including the symmetry energy at saturation density $J_{\rm sym,0}$, its slope $L_{\rm sym,0}$, and its curvature $K_{\rm sym,0}$, as well as higher-order terms $Q_{\rm sym,0}$ and $Z_{\rm sym,0}$. Furthermore, we detail various neutron star (NS) properties, such as the maximum NS mass $M_{\rm max}$, the radius corresponding to the maximum mass $R_{\rm max}$, the radius for a 1.4 solar mass NS $R_{1.4}$, and the tidal deformability for a 1.4 solar mass NS.  For reference $\Gamma_\rho$(DD2)$=7.2588$ and  $\Gamma_\rho$(DDMEX)$=7.2380$\label{tab:a1}}

\centering
\begin{tabular}{lcccccccccccccccc}
\hline 
\hline 
Model & $\Gamma_\rho$ & $a_\rho$ & $y_\rho$ & $M_{\rm dUrca}$ & ln $\mathcal{Z}_{\rm tot}$ & $\log \mathcal{B}_{M}^{i}$ & $J_{\rm sym,0}$ & $L_{\rm sym,0}$ & $K_{\rm sym,0}$ & $Q_{\rm sym,0}$ & $Z_{\rm sym,0}$ & $M_{\rm max}$ & $R_{\rm max}$ & $R_{1.4}$ & $\Lambda_{1.4}$ \\
\midrule
DD2B        & 7.489 & 0.927 & 0.637 & 2.351 & -1.99 & 0.00  & 32.50 & 49  & -48  & 762   & -9131  & 2.424 & 11.91 & 13.19 & 671 \\
DD2B-1.6    & 7.744 & 1.423 & 0.329 & 1.606 & -2.28 & -0.29 & 33.45 & 59  & 10   & 475   & -9019  & 2.438 & 12.01 & 13.38 & 739 \\
DD2B-1.8    & 7.675 & 1.350 & 0.397 & 1.808 & -2.12 & -0.13 & 33.19 & 53  & 7    & 558   & -10054 & 2.435 & 11.98 & 13.27 & 715 \\
DD2B-2.0    & 7.585 & 1.215 & 0.471 & 2.013 & -2.03 & -0.04 & 32.85 & 50  & -8   & 647   & -10241 & 2.432 & 11.95 & 13.22 & 697 \\
DDMEXB      & 7.302 & 1.049 & 0.495 & 2.239 & -2.29 & 0.00  & 32.49 & 58  & -14  & 575   & -9735  & 2.565 & 12.43 & 13.64 & 811 \\
DDMEXB-1.6  & 7.801 & 1.719 & 0.283 & 1.600 & -2.90 & -0.61 & 34.38 & 60  & 67   & 298   & -11309 & 2.574 & 12.49 & 13.62 & 846 \\
DDMEXB-1.8  & 7.509 & 1.567 & 0.320 & 1.796 & -2.40 & -0.11 & 33.26 & 59  & 48   & 369   & -11011 & 2.573 & 12.49 & 13.61 & 840 \\
DDMEXB-2.0  & 7.367 & 1.278 & 0.401 & 2.014 & -2.31 & -0.01 & 32.73 & 58  & 13   & 503   & -10502 & 2.570 & 12.46 & 13.63 & 825 \\
\bottomrule
\end{tabular}
\end{table*}

\bibliography{bibliosymdd1}
\bibliographystyle{apsrev4-1}
\end{document}